\documentclass[fleqn,10pt]{wlscirep}
\usepackage[utf8]{inputenc}
\usepackage[T1]{fontenc}
\usepackage{lineno}
\usepackage{amsmath,amssymb,amsfonts}
\usepackage{array}
\usepackage{wasysym}
\usepackage{textcomp}
\usepackage{stfloats}
\usepackage{url}
\usepackage{verbatim}
\usepackage{graphicx}
\usepackage{soul}
\usepackage{cite}
\usepackage{tikz}
\usepackage{hyperref}

\title{Diffusion backbone of temporal higher-order networks}
\author[1,*]{Shilun Zhang}
\author[2]{Alberto Ceria}
\author[1]{Huijuan Wang}
\affil[1]{Faculty of Electrical Engineering, Mathematics, and Computer Science, Delft University of Technology, Mekelweg 4, 2628 CD, Delft, The Netherlands}
\affil[2]{Institute of Advanced Computer Science, Leiden University, Einsteinweg 55, 2333 CC, Leiden, The Netherlands}
\affil[*]{S.Zhang-14@tudelft.nl}

\begin{abstract}

Temporal higher-order networks, where each hyperlink involving a group of nodes are activated or deactivated over time, are recently used to represent complex systems such as social contacts, interactions or collaborations that occur at specific times. Such networks are substrates for social contagion processes like the diffusion of information and opinions. In this work, we consider eight temporal higher-order networks derived from human face-to-face interactions in various contexts and the Susceptible-Infected threshold process on each of these networks: whenever a hyperlink is active and the number of infected nodes in the hyperlink exceeds a threshold $\Theta$, each susceptible node in the hyperlink is infected independently with probability $\beta$. The objective is to understand (1) the contribution of each hyperlink to the diffusion process, namely, the average number of nodes that are infected directly via the activation of the hyperlink when the diffusion starts from an arbitrary seed node, and (2) hyperlinks with what network properties tend to contribute more. Such understanding is crucial for further development of strategies to mitigate a diffusion process.
We first propose to construct the diffusion backbone. The backbone is a weighted higher-order network, where the weight of each hyperlink denotes the contribution of the hyperlink to a given diffusion process. Secondly, we find that the backbone, or the contribution of hyperlinks, is dependent on the parameters $\beta$ and $\Theta$ of the diffusion process, which is also supported by our theoretical analysis of the backbone when $\beta\rightarrow 0$. Thirdly, we systematically design centrality metrics, i.e., network properties, for hyperlinks in a temporal higher-order network and use each centrality metric to estimate the ranking of hyperlinks by the weight in the backbone. Finally, we find and explain why different centrality metrics can better estimate the contributions of hyperlinks for different parameters of the diffusion process. 
\end{abstract}
\begin{document}






\maketitle

\thispagestyle{empty}
\noindent

\section*{Introduction}
Complex networks serve as substrates for the diffusion of information, where a piece of information propagates along links connecting couples of nodes. Complex systems in nature and society are rarely static but exhibit time-varying network topologies \cite{holme2012temporal,holme2015modern,masuda2016guide}. Traditionally, such systems can be represented as temporal networks, where links between pairs of nodes are activated and deactivated over time. For example, a human physical contact network is usually experimentally recorded as a collection of time-resolved contacts, where a contact denotes an interaction between a pair of nodes at a specific timestamp. 
 
Prior works have revealed that properties of temporal networks such as the inter-event time distribution 
can affect the dynamics of processes unfolding on the temporal network, e.g., impact the speed of epidemic spreading or diffusion  \cite{karsai2011small, lambiotte2013burstiness, masuda2013temporal, rocha2013bursts, unicomb2021dynamics}, 
or impede the random walk explorations \cite{star2012random}.
The properties of nodes, node pairs, and subgraphs \cite{ciaperoni2020relevance} in temporal networks have been explored in order to identify which kind of nodes, node pairs, or subgraphs to activate (or block) to maximize (or to minimize) the spread of information (or epidemics) \cite{lee2012exploiting,starnini2013immunization,zhan2019information,zhang2022mitigate}. It was found that that vaccinating nodes with certain properties in a temporal network leads to a lower outbreak size when mitigating an epidemic spreading process on temporal networks \cite{lee2012exploiting,starnini2013immunization}. Ciaperoni et al. \cite{ciaperoni2020relevance} proposed to identify the subgraphs of a temporal network using a generalized $k$-core decomposition method and showed that the removal of temporal links belong to these subgraphs leads to large decrease in the final outbreak size of a spreading process.

Despite advances made in the past decade, studies of temporal networks have mostly focused on pairwise interactions, which fall short of representing a wide variety of real-world systems \cite{battiston2020networks,battiston2021physics,boccaletti2023structure,wang2024epidemic}. For example, individuals \cite{cencetti2021temporal,sekara2016fundamental} or animals \cite{iacopini2024not,musciotto2022beyond} may interact in a group of a size larger than two and the collaboration in a scientific paper may involve more than two researchers \cite{patania2017shape,abella2023unraveling}. Such higher-order (group) interactions can be represented as temporal higher-order networks, where a group or hyperlink is activated (when it has an interaction or contact) and deactivated over time. 
Previous studies have found shared properties of real-world temporal higher-order networks representing human physical interactions, such as the temporal correlation between hyperlinks in activity \cite{cencetti2021temporal,ceria2023temporal,iacopini2023temporal,gallo2024higher}.
It was also demonstrated that the properties of temporal higher-order networks influence the behavior of dynamical processes. For instance,
the duration of hyperlink activations was found to affect the onset of endemic state in epidemic spreading processes \cite{chowdhary2021simplicial}, while the time ordering of hyperlink activations impacts the consensus reached in nonlinear consensus dynamics \cite{neuhauser2021consensus}. This body of research has mainly revealed how global properties of the entire temporal higher-order network influence the behavior of dynamical processes.

However, the role or contribution of a hyperlink in a spreading process, e.g., the number of nodes that are infected directly via the activation of the hyperlink, on a temporal higher-order network remains unexplored. 
Recently, Zhan et al. \cite{zhan2019information} studied the contribution of pairwise links in a spreading process on a temporal pairwise network, finding that links that activate frequently earlier in time tend to contribute more when the infection probability is large. Contreras et al. \cite{contreras2024infection} investigated contagion processes on static higher-order network and found that the parameters of the contagion process affect the probability of a node being directly infected by another node. To understand which kind of hyperlinks contribute more to the diffusion process unfolding on temporal higher-order networks, new methods are required due to the new dynamics of the diffusion process.

In this paper, we aim to understand the role of hyperlinks and investigate which kind of hyperlinks, or hyperlinks with what properties, tend to contribute more to a diffusion process on the temporal higher-order network.
Empirical evidence has shown that in social phenomena, such as the diffusion of rumors or the adoption of norms and behaviors, contact with a single active neighbor is often insufficient to trigger adoption by an individual \cite{centola2010spread, karsai2014complex, monsted2017evidence}. Moreover, effects such as peer-pressure can occur as a consequence of the simultaneous exposition to many active members in a group gathering. As a result, a number of generalized models on higher-order networks have been proposed recently  to study social contagion, which is also referred as a spreading or diffusion process \cite{iacopini2019simplicial,de2020social,iacopini2022group}. In this work, we model the social contagion process by generalizing the Susceptible-Infected (SI) threshold spreading process, which is originally defined on a static higher-order network \cite{de2020social}, to a spreading process unfolding on a temporal higher-order network: initially, one seed node is infected while the other nodes are susceptible; when a hyperlink is active at any time, if the number of infected nodes within the hyperlink exceeds a threshold $\Theta$, each susceptible node in the hyperlink is independently infected with a probability $\beta$. The threshold $\Theta$ reflects how many exposures to infected nodes in a group interaction are required to trigger the infection of each susceptible node within the group. We propose to represent the contribution of each hyperlink to a diffusion process by constructing the diffusion backbone. The diffusion backbone is a static higher-order network that is the union of all hyperlinks that appear in at least one diffusion trajectory of the diffusion process, starting from an arbitrary seed node. Each hyperlink in the backbone is assigned a weight that reflects the average number of nodes directly infected through the activation of this hyperlink, over different choices of the seed node and realizations of the diffusion process. 

We construct the diffusion backbones for real-world temporal higher-order networks using the SI threshold processes with diverse parameters, $\beta$ and $\Theta$. Eight temporal higher-order networks derived from human face-to-face interactions in various contexts are considered. Firstly, we investigate how the infection probability $\beta$ and threshold $\Theta$ influence the constructed diffusion backbone. The backbone is shown to be dependent on the parameters of the diffusion process via both experiments and theoretical analysis of the backbone when $\beta\rightarrow 0$. Secondly, we explore which properties of a hyperlink tend to results in a large weight of the hyperlink in the diffusion backbone thus a significant contribution to the diffusion process. We propose centrality metrics, which encode local properties in the temporal higher-order network, for hyperlinks to estimate the ranking of hyperlinks by their contribution to the diffusion process. Each proposed metric is based only on the partial temporal higher-order network observed at the hyperlinks itself and its neighboring hyperlinks.  This allows efficient identification of hyperlinks that contribute more to the diffusion. For different regions of the process parameters, different (parameter-free) metrics perform the best, approaching the optimal performance of complex centrality metrics with control parameters. 
This is further explained through physical and theoretical interpretations.

Our findings elucidate the contributions of hyperlinks to a diffusion process. These findings could be insightful for the design of strategies to facilitate (or suppress)
the information diffusion on temporal higher-order networks via e.g., incentivizing the activation of selected groups.
\section*{Methods}
\label{sec:methods}
\subsection*{Temporal higher-order networks}
 A temporal (pairwise) network can be represented by $\mathcal{G}=(\mathcal{N}, \mathcal{C})$, where $\mathcal{N}$ is the set of nodes (or individuals), and $\mathcal{C}=\{(l(u,v), t)|u,v\in \mathcal{N}, t\in [1,T]\}$ is the set of events. Each event $(l(u, v), t)\in\mathcal{C}$ represents the pairwise interaction between node $u$ and $v$ occurring at discrete time $t$. 
 A temporal network $\mathcal{G}$ can be aggregated along the time dimension, giving the (time) aggregated network, denoted as $G=(\mathcal{N}, C)$. A link between the node pair $(u, v)$ exists in aggregated network $G$, i.e., $l(u,v) \in C$, if and only if there is at least one interaction in the temporal network between $u$ and $v$ during $[1, T]$. Each link $l(u,v)\in C$ in the aggregated network $G(\mathcal{N}, C)$ is indexed with an integer $j$, and the $j$-th link $l_j$ is associated with a weight $w_j$, which is the number of times that link $l_j$ has been activated in the temporal network.

However, people often gather in larger groups where more than two individuals interact simultaneously. The classic pairwise representation of temporal networks is limited in describing such group interactions, requiring the formalism of temporal higher-order networks.
A temporal higher-order network (or temporal hypergraph) is represented as $\mathcal{H}=(\mathcal{N}, \mathcal{E})$, where $\mathcal{E}=\{(h(u_1,...,u_d), t)|u_1,...,u_d\in \mathcal{N},t\in [1,T]\}$ is the set of higher-order interactions (events) involving an arbitrary number of nodes from the node set $\mathcal{N}$. Each higher-order event $(h(u_1,...,u_d), t)\in\mathcal{E}$ denotes a group interaction among the $d$ individual nodes at time $t$, where $h(u_1,...,u_d)=\{u_1,...,u_d\}$ denotes  an order-$d$ hyperlink among a set of the corresponding $d$ individuals. The size $d$ of the group is also referred to as the order of hyperlink $h(u_1,...,u_d)$. For example, the order of a dyadic and triadic hyperlink are $2$ and $3$ respectively. 
The higher-order aggregated network of a temporal higher-order network $\mathcal{H}$ is denoted as $H=(\mathcal{N}, E)$, where $E$ is the set of hyperlinks. Hyperlink $h(u_1,...,u_d)$ exists in $E$ if and only if hyperlink $h(u_1,...,u_d)$ has been activated at least once in the temporal higher-order network. Each hyperlink in the aggregated network $H$ is indexed with an integer $j$ and associated with a weight $w_j$, where $j\in[1,|E|]$ and $|E|$ is the total number of hyperlinks in $H$. The weight $w_j$ represents the number of times that hyperlink $h_j$ has been activated during $[1, T]$. We denote the activation of link $h_j$ in the temporal higher-order network by a time series $x_j(t)$, where $t \in [1, T]$ and $x_j(t) = 1$ if hyperlink $h_j$ is activated at time $t$, otherwise $x_j(t) = 0$. A temporal higher-order network $\mathcal{H}$ can thus be equivalently represented by its aggregated network $H$ with each link $h_j$ in $H$ further associated with its activity time series $x_j(t)$.
\subsection*{Diffusion process on temporal higher-order networks}
\label{ssec:model}

We consider a social contagion process on a temporal higher-order network, where each node is in one of two states at any time: infectious or susceptible. Initially, one seed node is infected while the other nodes are susceptible. Susceptible nodes can be infected through interactions with other infected nodes: when a hyperlink is active at any time, if the number of infected nodes within the hyperlink exceeds a threshold $\Theta$, each susceptible node within the hyperlink gets infected independently with a probability $\beta$; otherwise, it remains susceptible.
The traditional Susceptible-Infected (SI) process on a temporal pairwise network can be regards as a special case of the above diffusion process when all hyperlinks in $\mathcal{H}(\mathcal{N}, \mathcal{E})$ are dyadic (order-2), and the threshold is $\Theta=1$. When $\Theta=1$ and $\beta=1$, the diffusion process on a higher order temporal network becomes equivalent to the traditional SI process on the corresponding temporal pairwise network, where each higher-order interaction in the temporal higher-order network is treated as interactions between each pair of nodes within the hyperlink. We consider two cases in this paper: $\Theta=1$ and $\Theta=d-1$. When $\Theta=d-1$, the threshold for a hyperlink is dependent on the order $d$ of the hyperlink. Hence, hyperlinks of a higher order have a higher threshold.

Furthermore, the aforementioned diffusion model can be considered as a adjusted version of the model studied in Ref. \cite{de2020social}, with the following distinctions. We consider a discrete time Susceptible-Infectious (SI) diffusion process on temporal higher-order networks. In contrast, a continuous time Susceptible-Infectious-Susceptible (SIS) process on static higher-order networks has been studied in Ref. \cite{de2020social}. Our choice of discrete time process is because the underlying temporal higher-order networks evolve at discrete time. We start with the simple SI process instead of SIS process, which requires both network data of longer periods and better method to understand and identifies the steady state. 

\subsection*{Empirical datasets}
\label{ssec:datasets}
We apply our analysis to $8$ real-world human physical contact datasets from \textit{SocioPatterns}. These datasets contain collections of face-to-face interactions among individuals in various social contexts, including hospital (Hospital), primary school (Primaryschool2013), high school (Highschool2012, Highschool2013), workplace (Workplace2015), museum (Infectious), and conferences (HT2009, SFHH). The face-to-face interactions are recorded as pairwise contacts, where an interaction is stored when two individuals face each other at a distance of approximately $\lesssim 1.5$ meters over a $20$-second interval. Each original dataset naturally records only pairwise interactions, from which we deduce the corresponding temporal higher-order network via the common method already used in Ref. \cite{cencetti2021temporal,ceria2023temporal,gallo2024higher,iacopini2023temporal}. Specifically, at any time $t$, if there are $d(d-1)/2$ pairwise interactions between each nodes pair of a set of $d$ nodes, thus forming a clique, we promote these $d(d-1)/2$ pairwise interactions to an interaction of order $d$. For example, three temporal links $(l(a,b), t)$, $(l(b,c), t)$ and $(l(a,c), t)$ in the temporal network $\mathcal{G}(\mathcal{N}, \mathcal{C})$ at time $t$ are considered as a single temporal hyperlink $(h(a,b,c), t)$ in the corresponding temporal higher-order network $\mathcal{H}(\mathcal{N}, \mathcal{E})$. Since a clique of order $d$ contains all its sub-cliques of orders $d'<d$, only the maximal clique is promoted to a higher-order event. Furthermore, we preprocess the datasets by removing time steps without any interaction in the whole network and also excluding nodes that are not in the largest connected component of the higher-order aggregated network $H$. The basic statistics of each preprocessed dataset are presented in Table \ref{tab:datasets}.

\begin{table}[ht]
\centering
\begin{tabular}{@{}lllll@{}}
\toprule
Network        & $|\mathcal{N}|$ & $|E|$ & $|\mathcal{E}|$ & $T$ \\ \midrule
infectious     & 410  &  3350  & 14725  &  1393 \\
primaryschool  & 242  &  12704 & 106879  &  3101 \\
highschool2012 & 180  & 2645  & 42105  & 11274  \\
highschool2013 & 327  &  7818 & 172035  &  7376 \\
hospital       &  75 & 1825  & 27835  &  9454 \\
ht09           & 113 & 2434  & 19037  & 5247  \\
workplace15    & 217 & 4903 & 73823 & 18489\\
SFHH           & 403  & 10541  & 54306  &  3510 \\ \bottomrule
\end{tabular}
\caption{Statistics of real-world temporal higher-order networks after data processing. The number of nodes $|\mathcal{N}|$, the number of hyperlinks $|E|$, the number of higher-order events $|\mathcal{E}|$, and the number of time steps $T$ are shown.}
\label{tab:datasets}
\end{table}

\subsection*{Diffusion backbone}
\label{sec:analyticbackbone}
\begin{figure}[ht!]
    \centering
    \includegraphics[scale=0.8]{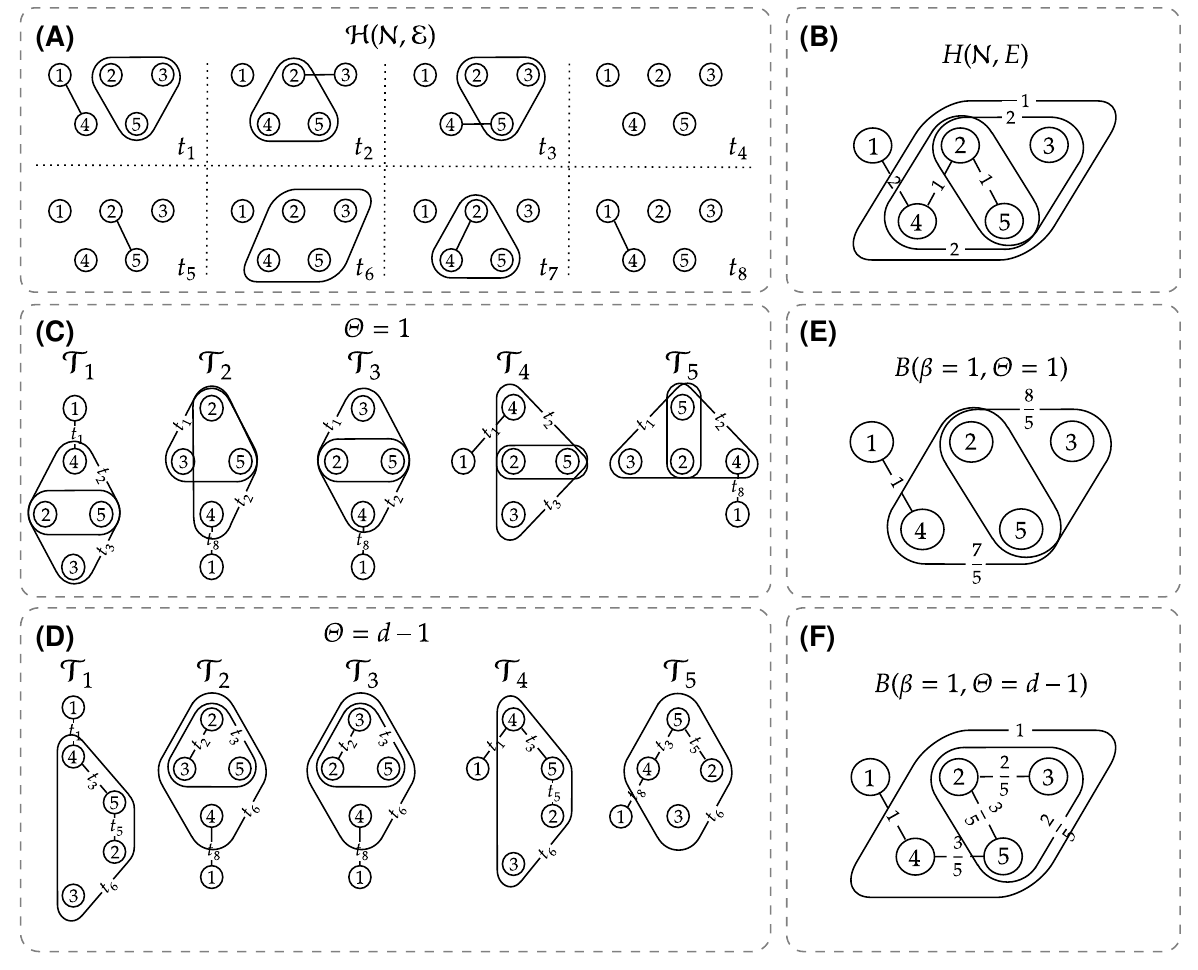}
    \caption{An illustration of the construction of the diffusion backbone of a temporal higher-order network when the infection probability of the diffusion model is $\beta=1.0$. (A) A temporal higher-order network $\mathcal{H}$ with $5$ nodes and $T=8$ time steps. (B) The time-aggregated higher-order network $H$, with a weight $w_j$ associated with each  hyperlink $j$. (C-D) Diffusion trajectory $\mathcal{T}_i (\beta=1)$ when $\Theta=1$ (panel C) and $\Theta=d-1$ (panel D) for each possible seed node $i$. (E-F) The diffusion backbone $B$ when $\Theta=1$(panel E) and when $\Theta=d-1$ (panel F).}
    \label{fig:illustration}
\end{figure}

Given a temporal higher-order network $\mathcal{H}(\mathcal{N}, \mathcal{E})$ and a diffusion process unfolding on $\mathcal{H}$, we quantify the contribution of a hyperlink (a group of nodes) to the diffusion process as the average number of nodes that are directly infected via the activation of the hyperlink. Specifically, we propose the following construction of a diffusion backbone to represent the contribution of each hyperlink to the diffusion process.
The diffusion backbone of $\mathcal{H}(\mathcal{N}, \mathcal{E})$ is a higher-order (static) network denoted as $B(\mathcal{N}, E_B)$. At time step $t=0$, one seed node is infected while all the other nodes are susceptible. For each seed node $i$, we construct a diffusion trajectory $\mathcal{T}_i(\Theta,\beta)$ as the union of the hyperlinks through which at least one susceptible node gets infected directly, during time period $[1, T]$. Each hyperlink $h_j$ in the trajectory $\mathcal{T}_i$ is associated with a weight $w^{\mathcal{T}_i}_j$ denoting the number of nodes in $h_j$ that are infected through the activation of hyperlink $h_j$ directly. The diffusion backbone is then defined as the union of all diffusion trajectories starting from each node in $\mathcal{N}$, i.e., $B(\Theta, \beta)=\cup_{i\in \mathcal{N}}\mathcal{T}_i(\Theta,\beta)$. Each hyperlink $h_j$ in $B(\Theta,\beta)$ is associated with a weight $w^B_j$, which is the average weights of the same hyperlink $h_j$ over all $|\mathcal{N}|$ diffusion trajectories, i.e., $w^B_j=\frac{1}{|\mathcal{N}|}\sum_{i\in\mathcal{N}}w^{\mathcal{T}_i}_j$. An illustration of the construction of the diffusion backbone in case of $\beta=1$ is shown in Figure \ref{fig:illustration}. In case of $0<\beta<1$, the diffusion process is stochastic, the diffusion backbone can be obtained by averaging over multiple independent realizations. In this work, we choose the number of realizations as $5\cdot 10^4$. The convergence of the backbone as the number of realizations increases is discussed in the Supplementary Information. The backbone $B(\mathcal{N}, E_B)$ encodes how many nodes on average are infected directly through each hyperlink when the seed node node is randomly selected. By definition, the total weights of all hyperlinks in the backbone $B(\beta,\Theta)$ plus one is equal to the average number $|\mathcal{N}|\rho$ of infected nodes during time $[1,T]$ per seed node, i.e., $\sum_{j\in E_B}w^B_j+1 = |\mathcal{N}|\rho$, since each infected node except the seed node leads to an increment of 1 in the weight of a hyperlink .

Now, we derived the backbone analytically for the limiting case when $\beta\rightarrow 0$. When $\Theta=1$, the diffusion backbone $B(\beta \rightarrow 0,\Theta=1)$ approaches the higher-order aggregated network $H(\mathcal{N},E)$ in topology, which can be explained as follows. Firstly, consider an arbitrary node $i$ as the seed node and one of its 1-hop neighbors $v$ in the higher-order aggregated network, i.e., $i$ and $v$ have at least a (group) interaction. The probability that the information diffuses from node $i$ to $v$ through an interaction that involves both nodes is $\beta$. Similarly, the total probability that $i$ infects $v$ via a
hyperlink $h_j$ 
is $\beta w_j$, where $w_j$ is the weight of hyperlink $h_j$ in the aggregated higher-order network, or equivalently, the total number of activations of $h_j$ in the temporal higher-order network. The total probability that the information diffuses from node $i$ to $v$ is the total weight of all hyperlinks that include both $i$ and $v$ in the aggregated higher-order network times $\beta$, thus of order $\beta$. Furthermore, consider a 2-hop neighbor $u$ of seed node $i$ in the aggregated higher-order network $H$,  i.e., node $u$ has interactions with at least one 1-hop neighbor of $i$ but has no interaction with node $i$. The probability that the information diffuses from node $i$ to a one hop neighbor of $i$, which spreads the information further to node $u$ is proportional to $\beta^2$, which is negligibly small compared to the probability for $i$ to infect a first-hop neighbor. Hence, the diffusion trajectory $\mathcal{T}_i$ starting from any seed node $i$ approaches the ego network of node $i$ in the aggregated higher-order network $H$, which comprises node $i$, its 1-hop neighbors and any hyperlink that includes $i$ and at least one of its 1-hop neighbors. Hence, the diffusion backbone $B(\beta\rightarrow 0, \Theta=1)$, which is the union of all $|\mathcal{N}|$ diffusion trajectories, has the same topology as the aggregated higher-order network $H$. The weight of an arbitrary hyperlink $h_j$ in the backbone is $w_j^B=\beta |h_j|(|h_j|-1)\cdot w_j \cdot \frac{1}{|\mathcal{N}|}$, because the activation of $h_j$ could spread the information only if a component node of $h_j$ is the seed and then each of the other $|h_j|-1$ component nodes could possible get infected. 

Consider the backbone when $\Theta=d-1$ and $\beta \rightarrow 0$. The same analysis applies to the weight $w_j^B$ of a dyadic hyperlink, thus the weight $w_j^B$ of a dyadic hyperlink is the same as that in the backbone $B(\beta \rightarrow 0,\Theta=1)$, i.e., $w_j^B=2\beta w_j\cdot \frac{1}{|\mathcal{N}|}$. The weight of an order-3 hyperlink $h_j$ in $B(\beta \rightarrow 0,\Theta=d-1)$ approaches $\frac{2}{|\mathcal{N}|}\beta^{2}\sum_{t=1}^{T} x_j(t)\cdot \left(\sum_{l\in \mathcal{L}^{sub}(j)} \sum_{\iota<t}x_l(\iota)\right )$ where the function $x_j(t)$ indicates whether the target hyperlink $h_j$ is activated at time $t$ (i.e., $x_j(t)= 1$) or not (i.e., $x_j(t)= 0$) and $\mathcal{L}^{sub}(j)$ includes all dyadic hyperlinks that share two common nodes with $h_j$. This second-order estimation can be explained as follows. Only when the seed node is a component node of $h_j$, could the activation of $h_j$ at any time $t$ infect a node with a probability of order $\beta^2$: the probability for a second node in $h_j$ to get infected before $t$ is $\beta$ times the total number of dyadic interactions the seed node has with the component nodes in $h_j$ before $t$. Taking into account the possibility that each component node of $h_j$ could be the seed node, the total probability that a second component node is infected before $t$ is $2\beta$ times the total number of activations $\sum_{l\in \mathcal{L}^{sub}(j)} \sum_{\iota<t}x_l(\iota)$ of all sub-links $\mathcal{L}^{sub}(j)$ of $h_j$ before $t$. If a second component node gets infected before $t$, the activation of $h_j$ could infect a third component node with probability $\beta$. If the seed node is outside $h_j$, the probability for the hyperlink to infect another node is negligibly small compared to that when the seed is a component node of $h_j$.

\subsection*{Observation time windows}
\begin{figure}[ht!]
    \centering
    \includegraphics[scale=1.1]{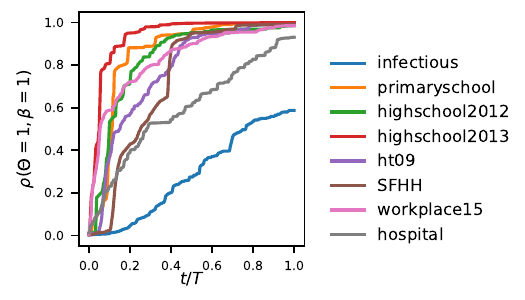}
    \caption{The average prevalence $\rho$ of the spreading process with threshold $\Theta=1$ and infection probability $\beta=1.0$ on each original temporal network over time $t$. The time steps are normalized by the total number of time steps $T$ of each dataset.}
    \label{fig:prevalence}
\end{figure}

Each real-world temporal higher-order network has a unique observation window $[1,T]$, determined by its measurement. To better understand the relationship between a hyperlink's network properties and its contribution to the diffusion process, should we consider the full observation window or only a portion of it as the dataset? To address this,
we examine the evolution of the average prevalence $\rho(t)$ over time, where $1\leq t \leq T$ and $T$ is the duration of original observation window, for each of the eight empirical temporal higher-order networks. This helps us understand events occurring at which time period may contribute to the diffusion, thus is relevant for the construction of the diffusion backbone. In case of $\Theta=1$ and $\beta=1.0$, the evolution of average prevalence $\rho$ at each timestamp is shown in Figure \ref{fig:prevalence}. In some networks like \textit{highschool2013}, the average prevalence increases rapidly in a short time and hardly anymore afterward, while in other networks like \textit{infectious}, the average prevalence increases continuously over time. This implies that in \textit{highschool2013}, the diffusion backbone $B(\Theta=1,\beta=1.0)$ is mainly determined by the interactions that occur in the early period, while in \textit{infectious}, all interactions during $[1, T]$ could contribute to the diffusion process. 
The above observation motivates us to consider various observation time windows from each data set when exploring the relation between properties of a hyperlink and its contribution in the diffusion process. Specifically, for each temporal higher-order network, we consider three time windows of different lengths, which are derived as follows. We examine the average prevalence $\rho(t)$ when $\Theta=1$ and $\beta=1.0$, where $1 \leq t \leq T$. We consider the following three observation windows, i.e.,  $[1,T_{p\%}]$, where $T_{p\%}$ is the earliest time when the average prevalence $\rho$ reaches $p\%\cdot \rho(T)$, i.e., $T_{p\%}=\min\{t: \rho(t)\geq p\%\cdot \rho(T)\}$, and $p\in\{30, 60, 90\}$. For each time window, we construct a temporal higher-order network that includes all the higher-order events occur during $[1,T_{p\%}]$. Eventually, a total of three temporal higher-order networks are derived from each dataset. In the following, we will present the results for the time window with the largest length, $[1, T_{90\%}]$ , and we refer to the Supplementary Information for results related to other time windows, which lead to qualitatively similar findings.
\section*{Results}
\label{sec:results}
In this section,  we will construct the diffusion backbone for each of the real-world temporal higher-order networks listed in Table \ref{tab:datasets} with its corresponding time window, e.g., $[1, T_{90\%}]$. We focus on two extreme cases for the threshold of the diffusion model: $\Theta=1$ and $\Theta=d-1$, with the infection probability $\beta$ ranging from $0.001$ to 1.0 on a logarithmic scale, i.e., $\beta\in[0.001, 1.0]$, to construct the backbones. 
To understand which kind of hyperlinks are used more in the diffusion process thereby acquire higher weights in the backbone under different parameters $\Theta$ and $\beta$, we will conduct the following analysis. Firstly, we will explore whether and how diffusion backbone changes with the two parameters $\beta$ and $\Theta$ of the diffusion model. 
This analysis will help understand whether different properties of hyperlinks should be used to estimate the ranking of hyperlinks by their weight in the backbone when the parameters of the diffusion process vary. Secondly, we will propose a set of hyperlink centrality metrics that capture diverse properties of a hyperlink in the temporal higher-order network. By examining the correlation between each proposed metric and the weight of a hyperlink in the backbone, we will uncover hyperlinks with what properties acquire higher weights in the diffusion backbone under different process parameters. In the following, we will present the results for \textit{SFHH} dataset with time window $[1, T_{90\%}]$, as the results for other datasets and different time windows show qualitatively similar trends (see Supplementary Information).

\subsection*{\textbf{Influence of process parameters on the backbone}}
\label{sec:parameter_influece_on_backbone}
In Section \ref{sec:analyticbackbone}, we have analytically derived the weighted backbone in the limit case of $\beta\rightarrow 0$. The backbone is shown to differ when $\Theta$ varies. Now, we explore how the backbone $B(\beta, \Theta)$ changes as the infection probability $\beta$ and threshold $\Theta$ vary.

\subsubsection*{\textbf{Total weight and number of links and in the backbone}}
\begin{figure}[h]
    \centering
    \includegraphics[scale=0.8]{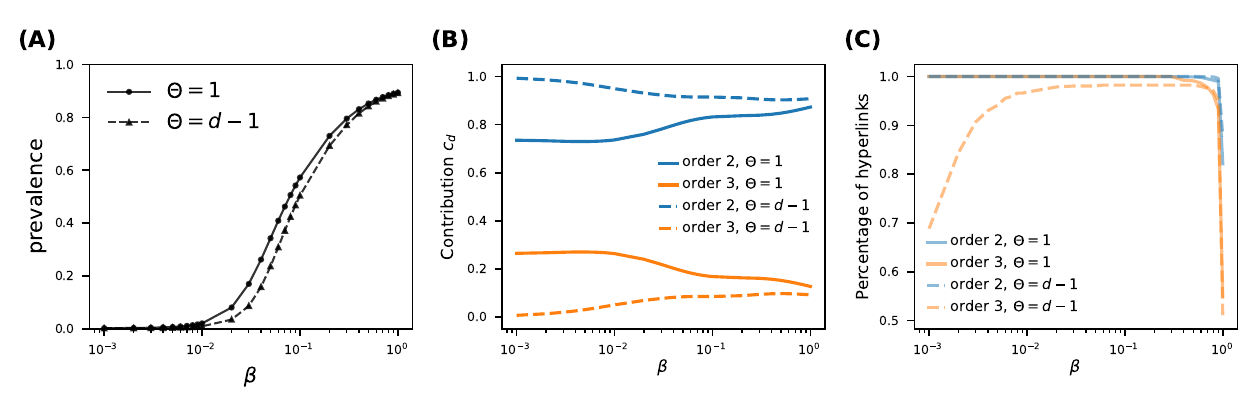}
    \caption{(A): The prevalence as a function of $\beta$ when $\Theta=1$ and $\Theta=d-1$ respectively. (B): The contribution of order-2 and order-2 hyperlinks as a function of $\beta$ when $\Theta=1$ and $\Theta=d-1$, respectively. (C): The percentage of order-2 and order-3 hyperlinks that appear in in backbone $B$ as a function of $\beta$ when $\Theta=1$ and $\Theta=d-1$. Results are shown for the \textit{SFHH} dataset.}
    \label{fig:influence_parameters}
\end{figure}

We first examine the prevalence of the diffusion process, which is equivalent to the total weights of all hyperlinks in the backbone plus one. In Figure \ref{fig:influence_parameters} (A), we show the prevalence as a function of the infection probability $\beta$. The wide range of prevalence under different $\beta$ and $\Theta$ shows that our constructed backbones span a broad dynamical space. Given the threshold $\Theta$, the prevalence increases with the infection probability $\beta$, suggesting that hyperlinks contribute more to the diffusion. The prevalence gets suppressed when $\Theta$ changes from 1 to $d-1$, because for the larger threshold $\Theta=d-1$, the diffusion through hyperlinks of higher orders ($d>2$) is impeded, which motivates us to explore the contribution of hyperlinks of different orders separately.



The relative contribution $c_d^B$ of order-$d$ hyperlinks to a diffusion process is reflected by the total weight of all order-$d$ hyperlinks in the backbone normalized by the total weight of all hyperlinks in the backbone. The relative contribution $c_d^B$ indicates the percentage of infected nodes that are infected through the activations of order-$d$ hyperlinks. Hence, $\sum_d c_d^B =1$. Since the number of hyperlinks of orders $d>3$ in each temporal higher-order network is small, we focus on hyperlinks of order 2 and order 3. Figure \ref{fig:influence_parameters} (B) shows the contribution $c_d^B$ of order 2 and order 3, respectively, under two distinct thresholds: $\Theta=1$ and $\Theta=d-1$, as $\beta$ increases. Generally, dyadic (order-2) hyperlinks exhibit the highest contribution, indicating that the majority of node infections occur through dyadic interactions. This is because order-2 interactions comprise the largest proportion of interactions in all the temporal higher-order networks and their threshold to spread information $\Theta=d-1=1$ is relatively low.
Given the same $\beta$, the relative contribution of order-3 links
when $\Theta=d-1$ is smaller than it is when $\Theta=1$, as expected. Let us consider the case when $\Theta=d-1$. When $\beta\rightarrow 0$, $c_2^B \rightarrow 1$ and the contribution of higher orders ($d>2$) approaches zero, which is consistent with our theoretical analysis (Section \ref{sec:methods}).
In this case, only few nodes get infected such that the number of infected nodes in a hyperlink of order $d>2$ can hardly reach $d-1$. As $\beta$ grows, more nodes get infected, which increases the chance that the information diffuses through triadic (higher-order) interactions. As a consequence, the contribution of triadic (dyadic) hyperlinks increases (decreases). When $\Theta=1$, the contribution of order-$3$ hyperlinks decreases as $\beta$ increases, because nodes within a triadic link could possibly get infected via interactions of dyadic links (large in number) before the activation of the triadic link, reducing the probability for a triadic link to diffuse the information.
Figure \ref{fig:influence_parameters} (C) shows the percentage of hyperlinks of a specific order $d$ in the aggregated higher-order network $H$ that appear in the backbone $B(\beta, \Theta)$. When $\Theta=1$ and $0<\beta<1$, each hyperlink in the temporal higher-order network $\mathcal{H}$ has a nonzero probability to appear in diffusion trajectories starting from every possible seed node. Thus, the diffusion backbone $B(\Theta=1, 0<\beta<1)$ contains all hyperlinks in the aggregated network $H$. While in the deterministic case of $\beta=1$, the diffusion from a source node to a target node follows the time-respecting paths that arrives at the target node the earliest in time. Hence, only hyperlinks that are present in such paths from one node to any other node are included in the diffusion backbone. This explains the drop in the fraction of hyperlinks that appears in the backbone as $\beta\rightarrow 1$, independent of $\Theta$. When $\Theta=d-1$ and $0<\beta<1$, each dyadic (order-2) hyperlink in the aggregated network appears in the backbone with a non-zero probability. However, not all order-3 hyperlinks necessarily appear in the backbone, because the condition $\Theta=d-1$ for an order-3 hyperlink to diffuse the information is possibly difficult to meet, especially when $\beta$ is small.
\subsubsection*{\textbf{Correlation between backbones under different process parameters}}

In this work, we aim to understand if a certain property of hyperlinks in the temporal higher-order network can be used to estimate the rankings of hyperlinks of the same order $d$ by the weight $w_j^B$ in the backbone. In the following, we study how the rankings of hyperlinks of a given order $d$ by the weight $w_j^B$ change with process parameters $\beta$ and $\Theta$. If the backbone changes with process parameters, different properties of hyperlinks maybe needed to identify hyperlinks with the highest backbone weight.

\begin{figure}[ht]
    \centering
    \includegraphics[scale=0.9]{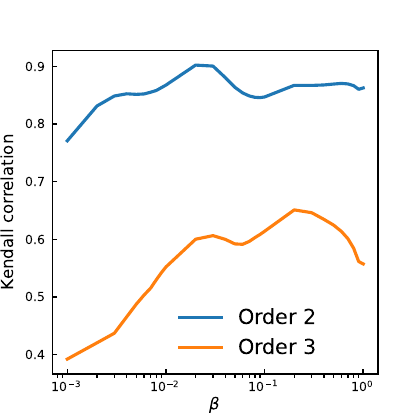}
    \caption{The Kendall correlation between $w_j^B(\Theta=1)$ and $w_j^B(\Theta=d-1)$ for order-2 hyperlinks and order-3 hyperlinks, in the \textit{SFHH} dataset.}
    \label{fig:corr_diff_theta}
\end{figure}

\begin{figure}[ht]
    \centering
    \includegraphics[scale=0.8]{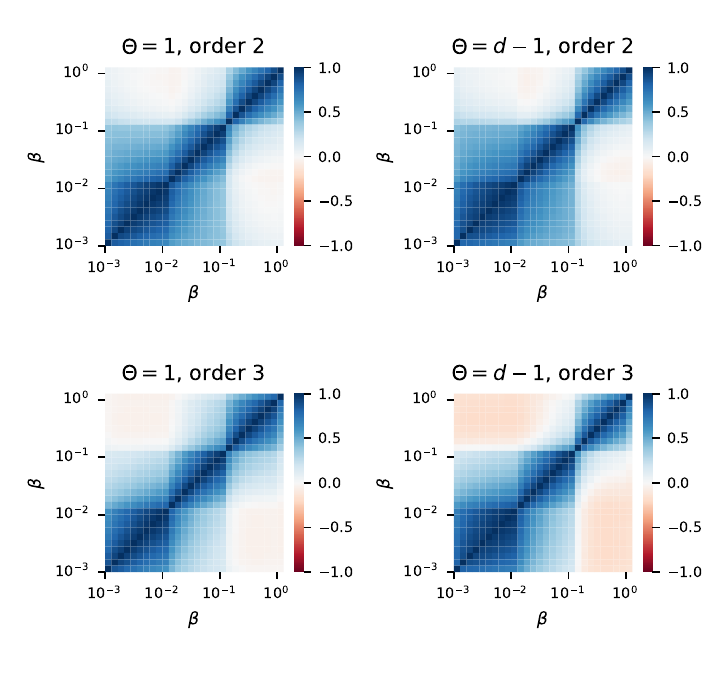}
    \caption{The Kendall correlation between the weight $w_j^B(\beta, \Theta)$ of an order-$d$ hyperlink obtained for two different infection probabilities, $\beta$, in the \textit{SFHH} dataset. Two left (right) panels correspond to the case of $\Theta=1$ ($\Theta=d-1$). Two upper (lower) panels show the Kendall correlation for hyperlinks of order $d=2$ (order $d=3$).}
    \label{fig:corr_diff_beta}
\end{figure}

Consider hyperlinks of order $d$ in the aggregated higher-order network $H$. We investigate the Kendall rank correlation between their weights in backbone $B(\beta, \Theta=1)$ and backbone $B(\beta, \Theta=d-1)$. Figure \ref{fig:corr_diff_theta} shows that the Kendall correlation for order-2 hyperlinks is in general higher than for order-3 hyperlinks, independent of $\beta$.
This suggests that different properties of hyperlinks may be needed to estimate the weights of order-3 hyperlinks in $B(\beta, \Theta=1)$ and in $B(\beta, \Theta=d-1)$, respectively. 
Furthermore, we examine the Kendall correlation between the weights of these hyperlinks in the backbones constructed with two different infection probabilities and the same threshold , which is shown in Figure \ref{fig:corr_diff_beta}. We find that when the difference between the two infection probabilities is small (large), the correlation is large (small). This suggests that backbones constructed under different infection probabilities differ in their link weights and different properties of hyperlinks may be required to estimate the ranking of hyperlinks in their backbone weight as $\beta$ varies.

\subsection*{\textbf{Centrality metrics for hyperlinks based on local temporal higher-order networks}}
\label{ssec:centrality}
Given a temporal higher-order network $\mathcal{H}$, which network property of a hyperlink is correlated with the weight of the hyperlink in the backbone $B(\beta, \Theta)$ under each parameter set $(\beta, \Theta)$? We will design different centrality metrics for a hyperlink, each reflecting a specific property of a hyperlink within its local temporal higher-order network. In the next subsection, we will investigate how well each metric can be used to estimate the ranking of hyperlinks based on their contribution to the diffusion process. 

The motivation for using local network information is threefold. Firstly, the information of local connections is more accessible than the global network information. Secondly, computing centrality metrics based on local connections is more efficient than those based on a larger set of connections.
Thirdly, local connections of a hyperlink could be more relevant than the rest of the temporal higher-order network, thus sufficient in estimating the weight of the hyperlink in the backbone, particularly when infection probability $\beta$ is relatively low. As discussed in Section \ref{sec:analyticbackbone}, when $\beta \rightarrow 0$ and $\Theta=1$, only 1-hop neighbors of the seed node could possibly get infected. Hence, the backbone weight of a hyperlink depends only on the number of times the hyperlink has been activated and the order of the hyperlink. As $\beta$ increases, the nodes that are further than 1- hop from the seed node may also get infected with a nonzero probability.
In this case, the activity (temporal connection) of neighboring hyperlinks of a target hyperlink could influence the weight of the target hyperlink in the backbone. For example, neighboring hyperlinks with a large number of activations may infect the component nodes of the target hyperlink, enabling the target hyperlink to meet its threshold condition and contribute to the spreading process.

We design local centrality metrics based on the activity of the target hyperlink itself and the activity of its neighboring hyperlinks. Since dyadic hyperlinks are more abundant than higher-order hyperlinks in each of the considered temporal higher-order networks, for an arbitrary target hyperlink $h_j$, we consider two different sets of neighboring dyadic hyperlinks respectively: all \textit{adjacent dyadic hyperlinks} that share at least one common node with the target $h_j$, denoted as set $\mathcal{L}^{adj}(j)$, and the \textit{dyadic sub-hyperlinks} that include all dyadic hyperlinks that share two common nodes with $h_j$, denoted as set $\mathcal{L}^{sub}(j)$. Any dyadic sub-hyperlink is also an adjacent dyadic hyperlink, so $\mathcal{L}^{sub}(j) \subseteq \mathcal{L}^{adj}(j)$. We propose firstly four centrality metrics capturing static or temporal properties of these two types of local neighborhoods, respectively.


\textbf{Static adjacent hyperlink based metric $\xi_j^{adj}$} of hyperlink $h_j$ is defined as:
\begin{align}
        \xi^{adj}_j(\alpha)= w_j\cdot \left(1+ \sum_{l\in \mathcal{L}^{adj}(j)}w_l \right )^{\alpha},
        \label{eqn:staticadj}
\end{align}

\textbf{Static sub-hyperlink based metric $\xi_j^{sub}$} is defined similarly as equation (\ref{eqn:staticadj}), except that the set of sub-hyperlinks $\mathcal{L}^{sub}(j)$ is considered instead of adjacent hyperlinks$\mathcal{L}^{adj}(j)$:
\begin{align}
        \xi^{sub}_j(\alpha)= w_j\cdot \left(1+ \sum_{l\in \mathcal{L}^{sub}(j)}w_l \right )^{\alpha}.
        \label{eqn:staticsub}
\end{align}

The static metrics $\xi_j^{adj}$ and $\xi_j^{sub}$ are determined by the number of activations of the target hyperlink, $w_j$, as well as the total number of activations of neighboring hyperlinks in set $\mathcal{L}^{adj}(j)$ and $\mathcal{L}^{sub}(j)$, respectively. The parameter $\alpha$ is the scaling parameter, which is a real constant 
and determines the contribution of neighboring hyperlinks to the metrics. Each proposed centrality metric is used to estimate the ranking of hyperlinks in backbone weight. Hence, using the logarithm
of the metric e.g., $\log(\xi^{adj})=\log w_j+\alpha \log\left(1+ \sum_{l\in \mathcal{L}^{adj}(j)}w_l \right)$, to predict the ranking is the same as using $\xi_j^{adj}$. This reveals also how $\alpha$ controls
the relative contribution of the adjacent links and the target link itself. These two metrics are motivated by the possibility that a hyperlink that has many activations in its neighborhood and in itself may contribute more to a diffusion process. 

\textbf{Temporal adjacent hyperlink based metric $\Xi_j^{adj}$} further considers the time ordering between the activations of the target hyperlink and the activations of hyperlinks in $\mathcal{L}^{adj}(j)$. We define the metric $\Xi^{adj}_j$ of a hyperlink $h_j$ as:
\begin{align}
        \Xi^{adj}_j(\alpha)=\sum_{t=1}^{T} x_j(t)\cdot \left(1+\sum_{l\in \mathcal{L}^{adj}(j)} \sum_{\iota<t}x_l(\iota)\right )^{\alpha},
        \label{eqn:temporaladj}
\end{align}

\textbf{Temporal sub-hyperlink based metric $\Xi_j^{sub}$} is defined in the same way as $\Xi^{adj}$, except that the contribution of neighboring hyperlinks in set $\mathcal{L}^{sub}(j)$ is considered. The metric $\Xi^{sub}_j$ of a hyperlink $h_j$ is defined as:
\begin{align}
        \Xi^{sub}_j(\alpha)=\sum_{t=1}^{T} x_j(t)\cdot \left(1+\sum_{l\in \mathcal{L}^{sub}(j)} \sum_{\iota<t}x_l(\iota)\right )^{\alpha}.
        \label{eqn:temporalsub}
\end{align}

In equation (\ref{eqn:temporaladj}) and (\ref{eqn:temporalsub}), the function $x_j(t)$ indicates whether the target hyperlink $h_j$ is activated at time $t$ (i.e., $x_j(t)= 1$) or not (i.e., $x_j(t)= 0$). Taking the metric $\Xi_j^{adj}$ as an example, it assigns a weight to each activation of the target hyperlink $h_j$. The weight for an activation at time $t$, given by $(1+\sum_{l\in \mathcal{L}^{adj}(j)} \sum_{\iota<t}x_l(\iota))^\alpha$, depends on the total number of activations of neighboring hyperlinks in set $\mathcal{L}^{adj}(j)$ that occurred before time $t$. Metric $\Xi_j^{adj}$ tends to be large if hyperlink $h_j$ has a large number of events (activations), and prior to each of its events, a large number of events have already occurred in adjacent hyperlinks.
Consider an order-3 target hyperlink, when $\Theta=d-1$ and $\beta$ is small, the target hyperlink could possibly infect its component nodes only when it is active (has an event) and two component nodes have already been infected at that time, which is more likely to happen when the neighboring hyperlinks have a large number of activations before. When $\beta$ is large, the large number of events occurring in the adjacent links before the activation of the target link could lead to the infection of all the component nodes of the target link, reducing the chance for the target hyperlink to infect its component nodes. Both scenarios motivate us to examine the number of events occurring in adjacent links before each event at the target hyperlink. Actually,  the design of $\Xi_j^{sub}$ is directly motivated by theoretical weight of an order-3 hyperlink in $B(\beta \rightarrow 0,\Theta=d-1)$ derived in Section \ref{sec:analyticbackbone}, i.e., $\frac{2}{|\mathcal{N}|}\beta^{2}\sum_{t=1}^{T} x_j(t)\cdot \left(\sum_{l\in \mathcal{L}^{sub}(j)} \sum_{\iota<t}x_l(\iota)\right )$. Hence, $\Xi_j^{sub}$ with $\alpha=1$ is supposed to estimate well the ranking of order-3 links in their weights in $B(\beta \rightarrow 0,\Theta=d-1)$.

When $\alpha= 0$, all metrics are equal to the weight $w_j$ of the target hyperlink in the higher-order aggregated network, i.e., $\xi^{adj}_j(0)=\xi^{sub}_j(0)=\Xi^{adj}_j(0)=\Xi^{sub}_j(0)=w_j$. When $\alpha>0$ ($\alpha<0$), the events in the neighborhood either $\mathcal{L}^{adj}(j)$ or $\mathcal{L}^{sub}(j)$, contribute positively (negatively) to each centrality metric. Order 2 hyperlinks have no neighboring sub-hyperlinks, i.e., the set $\mathcal{L}^{sub}(j)$ is empty, which means $\xi^{sub}_j(\alpha)=\Xi^{sub}_j(\alpha)=w_j$.

Furthermore, we consider a metric that has been proposed in \cite{zhan2019information} and used to estimate the diffusion backbone for temporal pairwise networks when $\beta=1$.

\textbf{Time-scaled weight $\Omega_j$} of hyperlink $h_j$ is defined as:
\begin{align}
    \Omega_j(\mu)=\sum_{t=1}^{T} x_j(t)\cdot t^\mu,
\end{align}
where $\mu$ is the scaling parameter. When $\mu=0$, the metric $\Omega_j$ equals the weight $w_j$ of hyperlink $h_j$ in the aggregated network. When $\mu<0$ ($\mu>0$), the metric $\Omega_j$ assigns higher (smaller) weights to activations that occur earlier in time. This metric is motivated by the possibility that a hyperlink that has activations that are large in number and early in time may contribute more to a diffusion process, especially when $\beta$ is relatively large. 

Finally, we combine the proposed temporal metrics, which capture the relatively time ordering of the activations of neighboring hyperlinks and the activations of the target hyperlink, with the metric time-scaled weight that captures the times of the activations of the target hyperlink itself. 

\textbf{Combined metrics $\Phi_j^{adj}$ and $\Phi_j^{sub}$} combine the time-scaled weight $\Omega_j$ with metric $\Xi_j^{adj}$ and $\Xi_j^{sub}$, respectively. 
\begin{align}
    \Phi_j^{adj}(\mu,\alpha)= \sum_{t=1}^{T} x_j(t)\cdot t^{\mu}\cdot\left(1+\sum_{l\in \mathcal{L}^{adj}(j)} \sum_{\iota<t}x_l(\iota)\right )^{\alpha},
\end{align}
\begin{align}
    \Phi_j^{sub}(\mu,\alpha)= \sum_{t=1}^{T} x_j(t)\cdot t^\mu\cdot\left(1+\sum_{l\in \mathcal{L}^{sub}(j)} \sum_{\iota<t}x_l(\iota)\right )^{\alpha},
\end{align}
where the real constants $\mu_j$ and $\alpha$ are two scaling parameters. We have $\Phi_j^{adj}(\mu=0,\alpha)=\Xi_j^{adj}$, $\Phi_j^{sub}(\mu=0,\alpha)=\Xi_j^{sub}$, $\Phi_j^{adj}(\mu,\alpha=0)=\Phi_j^{sub}(\mu,\alpha=0)=\Omega_j(\mu)$.

\subsection*{\textbf{Estimating hyperlink weight $w_j^B$ using local centrality metrics}}
\subsubsection*{\textbf{New metrics with a single parameter $\alpha$}}

\begin{figure}[h]
    \centering
    \includegraphics[scale=0.7]{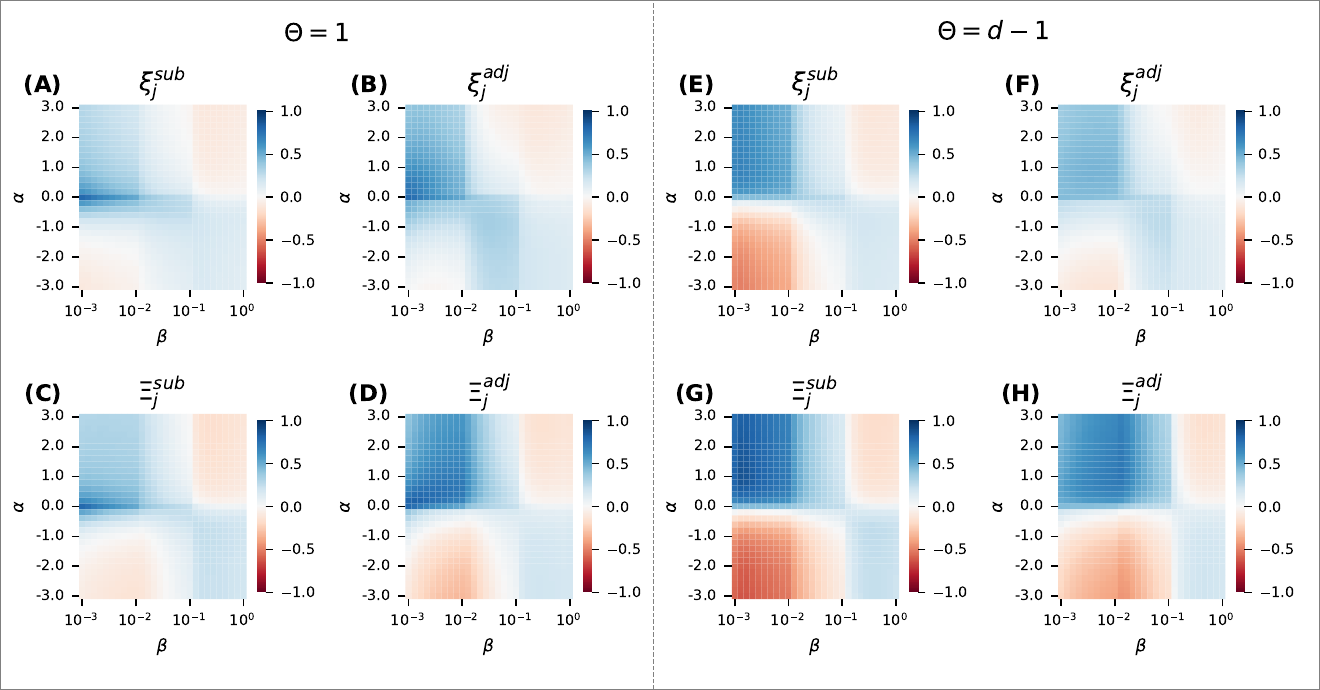}
    \caption{The Kendall correlation between centrality metric $\xi_j^{adj}(\alpha)$, $\xi_j^{sub}(\alpha)$, $\Xi_j^{adj}(\alpha)$ or $\Xi_j^{sub}(\alpha)$ and the weight $w_j^B$ of an order-3 hyperlink in the backbone $B$, as a function of the infection probability $\beta$ and the scaling parameter $\alpha$, for $\Theta=1$ (left panels, A-D) and $\Theta=d-1$ (right panels, E-H). The results are shown for the \textit{SFHH} dataset. }
    \label{fig:heatmap_order3_SFHH}
\end{figure}
\begin{figure}[h]
    \centering
    \includegraphics[scale=0.9]{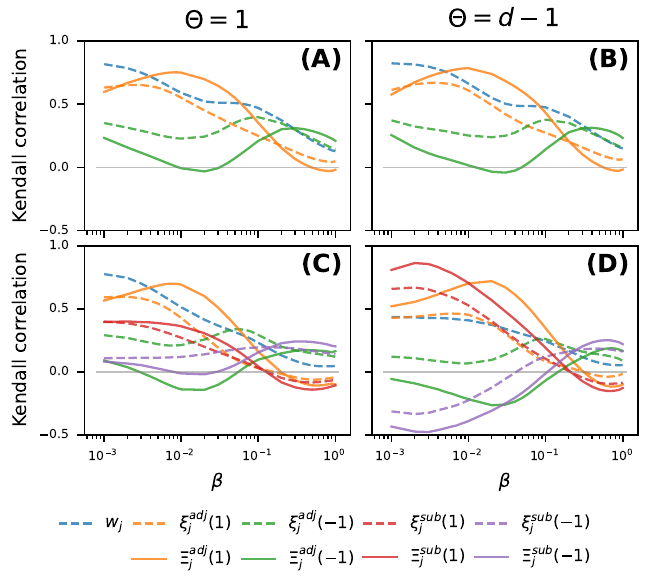}
    \caption{The Kendall correlation between a centrality metric with the scaling parameter $\alpha\in\{0, 1, -1\}$ and the weight $w_j^B$ of a hyperlink in the backbone $B$, as a function of infection probability $\beta$, in the \textit{SFHH} dataset. Results are shown for dyadic hyperlinks (A-B) and triadic hyperlinks (C-D). Two columns corresponds to $\Theta=1$ and $\Theta=d-1$, respectively.}
    \label{fig:SFHH_corr}
\end{figure}

We examine the performance of each centrality metric with the scaling parameters $\alpha\in [-3, 3]$ and $\mu\in [-3, 3]$ in predicting the ranking of hyperlinks of a specific order $d$ by the weights $w_j^B$. It is challenging to pre-select the parameter(s) of a metric when performing such a prediction task, especially when the number of parameters is large.
Hence, we first evaluate the performance of the four new metrics with a single parameter $\alpha$: $\xi_j^{adj}(\alpha)$, $\xi_j^{sub}(\alpha)$, $\Xi_j^{adj}(\alpha)$, $\Xi_j^{sub}(\alpha)$, our main focus. The objectives are to (1) understand how the scaling parameter $\alpha$ affects their performance and whether certain values of the scaling parameter lead to the highest performance of a metric across all networks, and (2) compare the performance of these four metrics. Later in Section \ref{ssec:optimal}, we will compare the performance of these four metrics with the other three metrics we have introduced in Section \ref{ssec:centrality}, time-scaled weight from literature and two combined metrics. The objective is to explore whether these four metrics with one parameter or with evident choice of the parameter could perform close to the combined metrics with two scaling parameters. 

Figure \ref{fig:heatmap_order3_SFHH} (A-H) show the Kendall rank correlation between each of the four new metrics and the weight $w_j^B$ of an order-3 hyperlink as a function of the scaling parameter $\alpha$ and the infection probability $\beta$, when $\Theta=1$ (panel (A-D)) and $\Theta=d-1$ (panel (E-H)). 
A general observation is that when $\beta$ is relatively small (large), each metric with a positive (negative) $\alpha$ of a hyperlink is positively correlated with the weight of the hyperlink in the backbone. This is also observed for order-2 hyperlinks. Indeed, when $\beta$ is small, a large number of activations of neighboring hyperlinks could increase the chance that nodes in the target hyperlink get infected, which increases the chance that activations of the target hyperlink afterwards could infect other nodes. However, when $\beta$ is sufficiently large, the chance that all component nodes in the target hyperlink get infected through the activations of neighboring hyperlinks may increase, which suppresses the chance that the activation of the target hyperlink afterwards could infect other nodes.

Given the parameters $\beta$ and $\Theta$, the best prediction performance of each metric is roughly achieved when $\alpha =0, 1$ or $-1$ in each real-world network, which is further shown by the comparison of the performance of each metric with $\alpha =0, 1$, $-1$ and its best performance across $\alpha\in [-3,3]$ in Figure S6 in Supplementary Information.
Hence, we will focus metrics $w_j$, $\xi_j^{adj}(1)$, $\xi_j^{adj}(-1)$, $\Xi_j^{adj}(1)$, $\Xi_j^{adj}(-1)$, $\xi_j^{sub}(1)$, $\xi_j^{sub}(-1)$, $\Xi_j^{sub}(1)$, $\Xi_j^{sub}(-1)$, where the parameter has been selected. Since order-2 hyperlinks have no sub-hyperlinks, which makes $\xi^{sub}_j(\alpha)=\Xi^{sub}_j(\alpha)=w_j$ according to the definitions, only metrics, $w_j$, $\xi_j^{adj}(1)$, $\xi_j^{adj}(-1)$, $\Xi_j^{adj}(1)$, $\Xi_j^{adj}(-1)$, will be considered for order-2 hyperlinks. 

Figure \ref{fig:SFHH_corr} shows the Kendall correlation between each metric with $\alpha \in \{0, 1, -1\}$ with the weight $w_j^B$ of a hyperlink as a function of $\beta$, for order-2 and order-3 hyperlinks respectively. Generally, the best-performing centrality metric varies depending on the process parameters $\Theta$ and $\beta$, because the backbone is dependent on the process parameters. Consider the region of $\beta\rightarrow 0$. Metric $\Xi^{sub}_j(1)$ exhibits the highest correlation with the weight $w_j^B$ of order-3 hyperlinks in backbone $B(\beta \rightarrow 0,\Theta=d-1)$, as shown in Figure \ref{fig:heatmap_order3_SFHH} (D), while metric $w_j$ is the best-performing metric to estimate the weight $w_j^B$ of order-2 hyperlinks in backbone $B(\beta \rightarrow 0,\Theta=d-1)$ (Figure \ref{fig:SFHH_corr} (B)) and to estimate the weight $w_j^B$ of hyperlinks of both order 2 (Figure \ref{fig:SFHH_corr} (A)) and order 3 (Figure \ref{fig:SFHH_corr} (C)) in backbone $B(\beta \rightarrow 0,\Theta=1)$. These two observations are in line with the backbone $B(\beta \rightarrow 0, \Theta)$ derived analytically in Section \ref{sec:analyticbackbone}.

As $\beta$ increases but is still small ($<10^{-1}$), metric $\Xi^{adj}_j(1)$ outperforms the other metrics. In this range of $\beta$, nodes that are more than 1 hop away from the seed could be infected with a non-zero probability. A larger number of activations of the adjacent hyperlinks could lead to a higher probability of infection of the component nodes of the target hyperlink,  which makes the target hyperlink more likely to meet the threshold to infect other component node(s) when activated.

When $\beta$ is large ($>10^{-1}$), our temporal metrics with $\alpha=-1$ tend to perform the best. However, their performance is still relatively low, i.e., the correlation with backbone weight is evidently lower than 0.5. This suggests that when $\beta$ is large, the diffusion becomes global, local network information alone may be insufficient to well predict the contribution of a hyperlink to the diffusion process. 

In general, when the infection probability $\beta$ is not large ($\beta < 10^{-1}$), either $\Xi_j^{adj}(1)$, $\Xi_j^{sub}(1)$ or $w_j$ performs best, depending on the process parameters and the order of hyperlinks. These observations are qualitatively similar in other datasets and different observation time windows, though the specific range of $\beta$ for the a metric to perform the best may vary (see Supplementary Information).

\begin{figure}[h]
    \centering
    \includegraphics[scale=0.8]{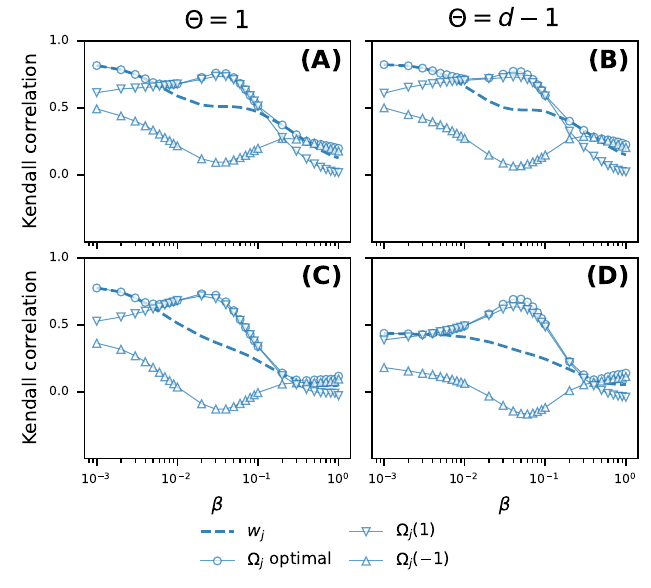}
    \caption{The Kendall correlation between the time-scaled weight $\Omega_j(\mu)$ with $\mu\in\{0, 1, -1\}$ and the weight $w_j^B (\beta,\Theta)$ of a hyperlink in the backbone, in \textit{SFHH} dataset. This is compared with the optimal Kendall correlation (circles) of $\Omega_j(\mu)$ achieved by all possible choices of $\mu\in[-3,3]$. Results are shown for dyadic hyperlinks (A-B) and triadic hyperlinks (C-D). Two columns corresponds to $\Theta=1$ and $\Theta=d-1$, respectively.}
    \label{fig:Omega_compare}
\end{figure}
\begin{figure}[h]
    \centering
    \includegraphics[scale=0.8]{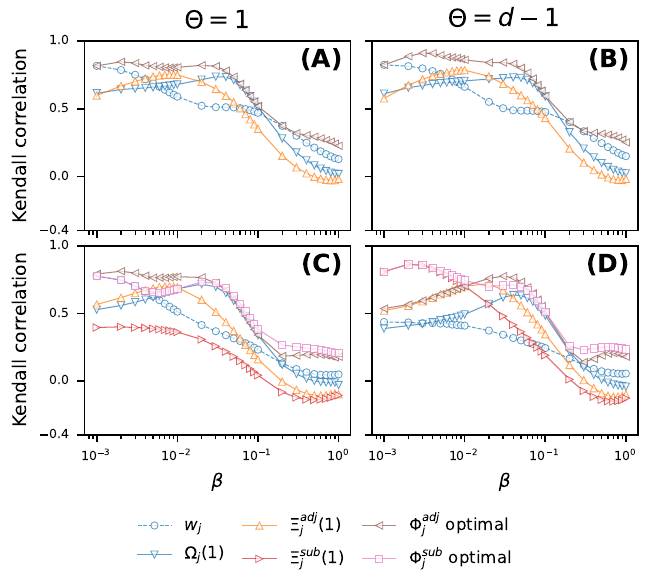}
    \caption{The Kendall rank correlation between each of the four metrics $w_j$, $\Omega_j(1)$, $\Xi_j^{adj}$ and $\Xi_j^{sub}$ and the weight $w_j^B$ of a hyperlink in the backbone, in comparison to the optimal Kendall correlation between each combine metric and the weight $w_j^B$, in \textit{SFHH} dataset. Results are shown for dyadic hyperlinks (A-B) and triadic hyperlinks (C-D). Two columns corresponds to $\Theta=1$ and $\Theta=d-1$, respectively.}
    \label{fig:combinedmetrics}
\end{figure}
\subsection*{Comparison with other temporal centrality metrics}
\label{ssec:optimal}
We further compare the performance of $\Xi_j^{adj}(1)$, $\Xi_j^{sub}(1)$ and $w_j$ with the other three metrics: the time-scaled weight $\Omega_j(\mu)$ and the two combined metrics $\Phi_j^{adj}(\mu,\alpha)$, $\Phi_j^{sub}(\mu,\alpha)$, in predicting the ranking of hyperlinks of a given order by their weights in the backbone $B(\beta,\Theta)$. 

We first investigate whether the metric $\Omega_j(\mu)$ with a specific choice of the scaling parameter $\mu$ tends to lead to the highest prediction performance across all networks. As shown in Figure \ref{fig:Omega_compare},
when $\beta$ is small ($\beta<10^{-2}$), the optimal/highest Kendall correlation between metric $\Omega_j(\mu)$ and the weight $w_j^B$, is achieved approximately by $w_j=\Omega_j(0)$. When $\beta$ increases but is still smaller than $10^{-1}$, the performance of $\Omega_j(1)$ is close to the optimal Kendall correlation, indicating that hyperlinks that activate frequently later in time tend to contribute more. As $\beta$ increases further ($\beta>10^{-1}$), the optimal Kendall correlation is approximately achieved by metric $\Omega_j(-1)$, suggesting that hyperlinks that activate earlier in time tend to contribute more to the diffusion process when $\beta$ is large. Next, we will compare so far the best performing metrics $\Xi_j^{adj}(1)$, $\Xi_j^{sub}(1)$, $w_j$ and $\Omega_j(1)$, which do not require parameter calibration, with the two combined metrics in estimating the ranking of hyperlinks of a given order by their weights in the backbone $B(\beta,\Theta)$. 
  
Given the infection probability $\beta$ and the threshold $\Theta$, we perform a grid search for the scaling parameters $\mu$ and $\alpha$ within the range $[-3.0, 3.0]$ to find the highest Kendall correlation of each combined metric when estimating the ranking of order-2 and order-3 hyperlinks, respectively. The optimal performance of a combined metric, say $\Phi_j^{adj}$, is the upper bound for the performance of $\Xi_j^{adj}$, $w_j$ and the time-scaled weight $\Omega_j$, which are special cases of the combined metric.
As shown in Figure \ref{fig:combinedmetrics}, the best performance achieved by metrics $\Xi_j^{adj}(1)$, $\Xi_j^{sub}(1)$, $w_j$ and $\Omega_j(1)$ is close to the optimal performance of the two combined metrics when $\beta < 10^{-1}$. We compare further the performance of the four metrics without the need for parameter calibration: $\Xi_j^{adj}(1)$, $\Xi_j^{sub}(1)$, $w_j$ and $\Omega_j(1)$. When $\beta\rightarrow 0$, either metric $w_j$ or $\Xi^{sub}_j(1)$ performs the best depending on $\Theta$ and the order of hyperlinks under estimation, which is in line with our analytical analysis. As $\beta$ increases ($\beta\approx 10^{-2}$), $\Xi_j^{adj}(1)$ outperforms. A large number of activations of the adjacent hyperlinks could cause the infection of component nodes in a target hyperlink thus fulfilling the threshold condition for the target to infect other nodes. As $\beta$ increases further but still within the range of $\beta < 10^{-1}$, $\Omega_j(1)$ performs the best. A hyperlink with a large number of activations that occur relatively late in time tend to contribute more the spreading process. Activations that occur late, when more nodes are infected thus the threshold condition is likely met, tend to contribute effectively to the spreading process. 
In summary, for different ranges of the infection rate $\beta$, different parameter free centrality metrics estimate the backbone weights the best, close to the optimal performance of combined metrics achieved by parameter searching.

\section*{Conclusion and future work}
\label{sec:discussion}
In this paper, the contribution of a hyperlink in a higher-order temporal network to a diffusion process is defined as the average number of nodes infected via the activation of the hyperlink. We explored which properties of a hyperlink in the higher-order temporal network lead to a high contribution. A generalized Susceptible-Infected threshold process with infection probability $\beta$ and threshold $\Theta$ is considered on eight real-world higher-order temporal networks derived from human face-to-face contacts in various contexts. 
Firstly, we proposed the construction of the diffusion backbone $B$ where the weight of each hyperlink equals the the contribution of the hyperlink to the diffusion process starting from a random seed node. In the limiting case when $\beta\rightarrow 0$, the backbone weight of a hyperlink was derived analytically based on local temporal connections around the hyperlink. Secondly, we illustrated the dependency of the backbone $B(\beta, \Theta)$ on the two process parameters $\beta$ and $\Theta$. This is also evidenced by the different backbones derived when $\Theta$ varies and $\beta\rightarrow 0$. To explore which properties of hyperlinks are associated with high contributions to the diffusion process, we designed centrality metrics for a hyperlink to estimate the ranking of hyperlinks by their weights in backbone $B(\beta, \Theta)$. Each proposed metric is defined based on the activity of the target hyperlink and the activity of its neighboring hyperlinks in the higher-order temporal network. Different metrics are shown to predict the best for different parameter sets $(\beta, \Theta)$ of the process, approaching the optimal performance of combined metrics achieved via parameter searching.
The reason why certain properties of a hyperlink result in a high contribution to the process is further explained.

There are several limitations in this work that call for further exploration. Firstly, we only considered the Susceptible-Infected threshold process as the diffusion model on higher-order temporal networks and focused only on two extreme cases for the threshold $\Theta$. A more comprehensive investigation of diffusion models is needed. 
Secondly, it is interesting to explore other types of higher-order temporal networks, such as scientific collaborations, which may have different properties than the human interaction networks considered in this work. Thirdly, the backbone constructed captures the contribution of each hyperlink. This definition can be extended to study the contribution of each node. Finally, it is promising to investigate how to use the backbone or our proposed centrality metrics that well estimate the backbone to mitigate the dynamic process on the network via the blocking of selected hyperlinks.


\pagebreak
\makeatletter
\renewcommand \thesection{S\@arabic\c@section}
\renewcommand\thetable{S\@arabic\c@table}
\renewcommand \thefigure{S\@arabic\c@figure}
\makeatother

\begin{center}
\textbf{\sffamily\bfseries\LARGE Supplementary information: diffusion backbone of temporal higher-order networks}
\end{center}

\maketitle



%


\section{Statistics of empirical datasets}
\begin{figure*}[h]
\centering
\includegraphics[scale=0.7]{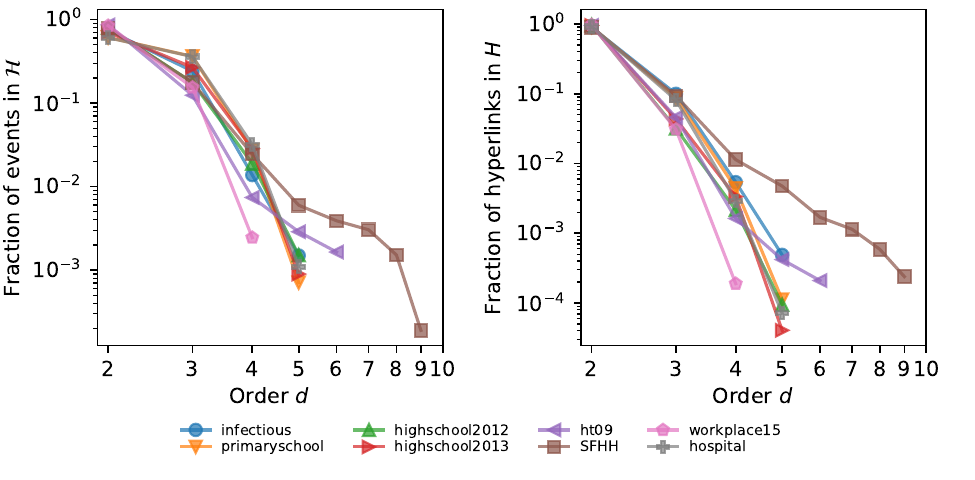}
\caption{The fraction of events (hyperlinks) that are of order $d$ in each real-world temporal higher-order network with the original observation time window $[1,T]$ is shown in the left (right) panel.}
\label{sfig:order_statistics}
\end{figure*}

\begin{table}[ht]
\centering
\begin{tabular}{@{}cccc@{}}
\toprule
       Dataset & $T_{30\%}$ & $T_{60\%}$ & $T_{90\%}$ \\ \midrule
       infectious & 519 & 784 & 1104 \\ 
       primaryschool & 287 & 359 & 994  \\ 
       highschool2012 & 916 & 1477 & 3664 \\ 
       highschool2013 & 195 & 395  & 1252 \\ 
       hospital & 1266 & 3942  & 7702 \\ 
       ht09 & 437 & 1154 & 2361 \\ 
       workplace & 574 & 2133 & 8773 \\ 
       SFHH & 483 & 1124 & 1421 \\ \bottomrule
    \end{tabular}
\caption{The number of time steps in the observation time windows $[1,T_{p\%}]$ that we choose for each empirical dataset.}
\label{tab:time_windows}
\end{table}

\section{Convergence of diffusion backbone}
We explore whether $R=50000$ realizations of the diffusion process starting from each seed node is sufficient to obtain a representative backbone for $0 < \beta < 1$. Given the infection probability $\beta$ and the threshold $\Theta$, we construct the diffusion backbones $B(\beta, \Theta)$ resulting from different numbers $R$ of independent realizations. Then, we
measure the Kendall correlation between the backbones in weight resulting from $R$ realizations and $50000$ realizations, for order-2 and order-3 hyperlinks, respectively. 

\begin{figure*}[h]
\centering
\includegraphics[scale=0.55]{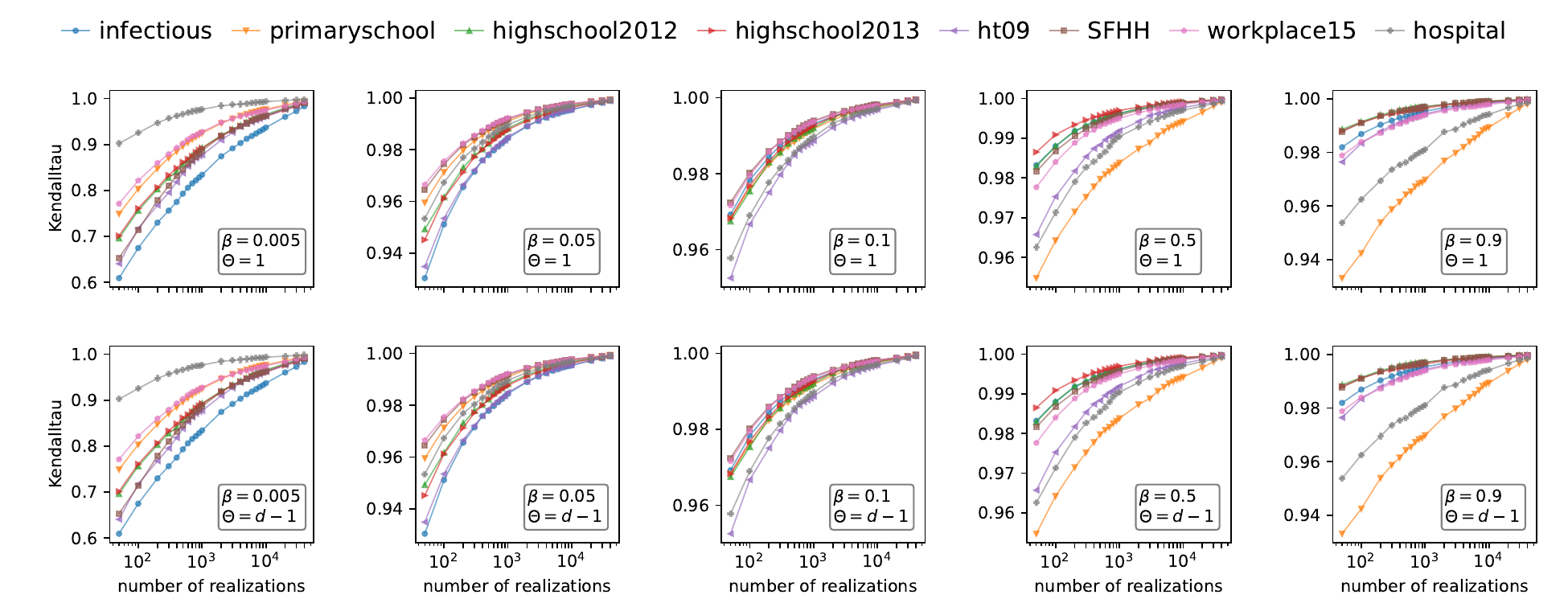}
\caption{The Kendall correlation between the weights in the backbones resulting from $R$ realizations and $50000$ realizations, as a function of $R$ for order-2 hyperlinks. This is shown for each empirical dataset with the observation time window $[1, T_{90\%}]$.}
\label{sfig:convergence_all_order2}
\end{figure*}

\begin{figure*}[h]
\centering
\includegraphics[scale=0.55]{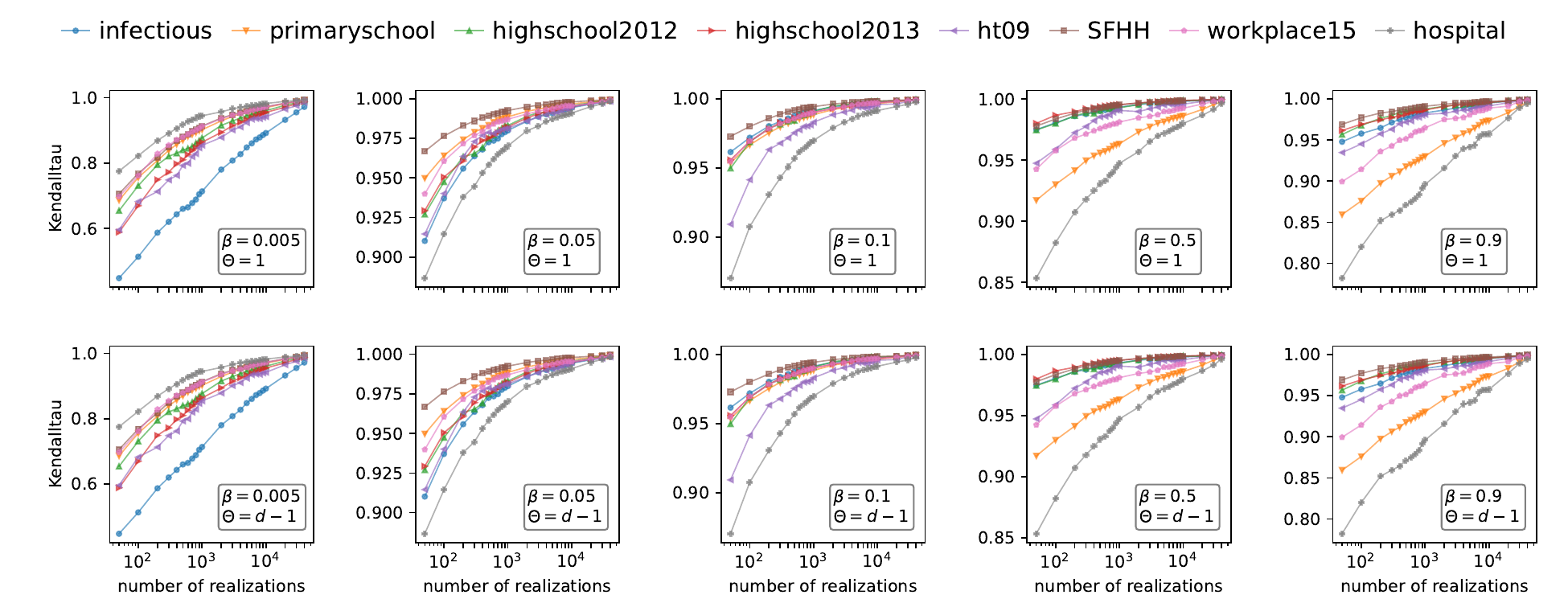}
\caption{The Kendall correlation between the weights in the backbones resulting from $R$ realizations and $50000$ realizations, as a function of $R$ for order-3 hyperlinks. This is shown for each empirical dataset with the observation time window $[1, T_{90\%}]$.}
\label{sfig:convergence_all_order3}
\end{figure*}

\begin{figure*}[h]
\centering
\includegraphics[scale=0.55]{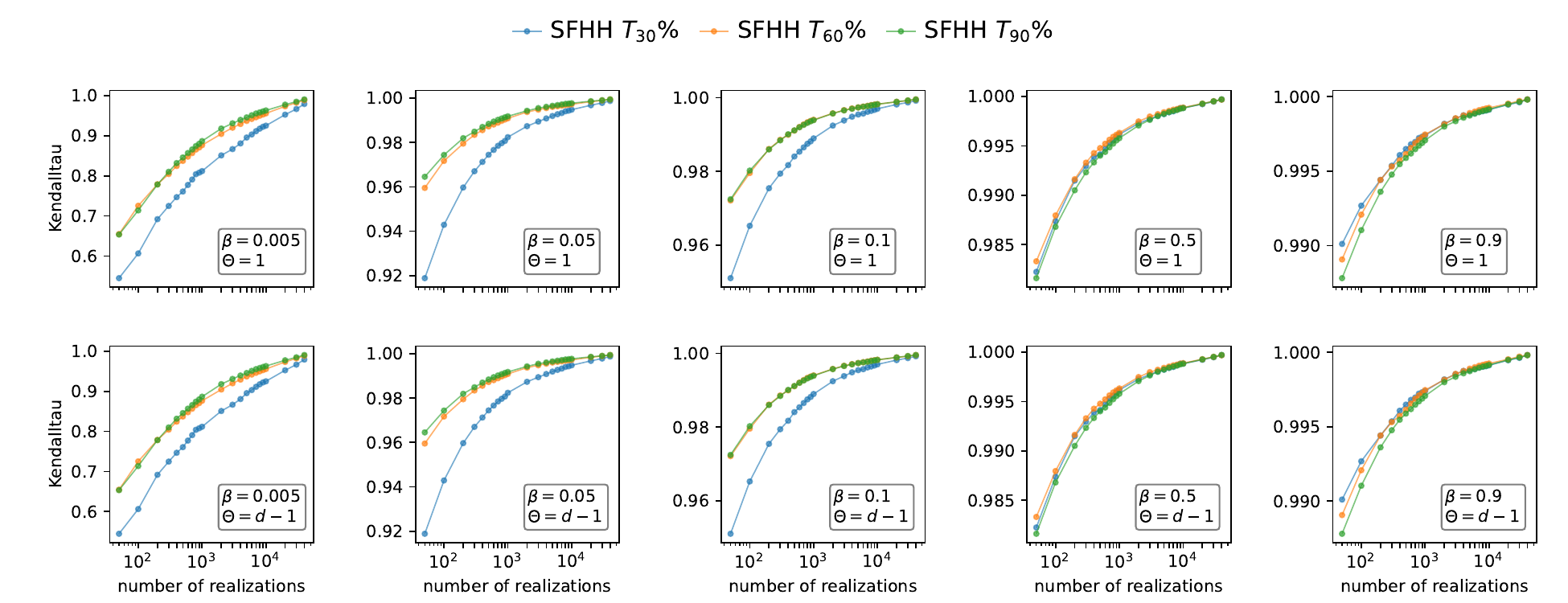}
\caption{The Kendall correlation between the weights in the backbones resulting from $R$ realizations and $50000$ realizations, as a function of $R$ for order-2 hyperlinks, in \textit{SFHH} dataset with different observation time windows.}
\label{sfig:convergence_SFHH_order2}
\end{figure*}

\begin{figure*}[h]
\centering
\includegraphics[scale=0.55]{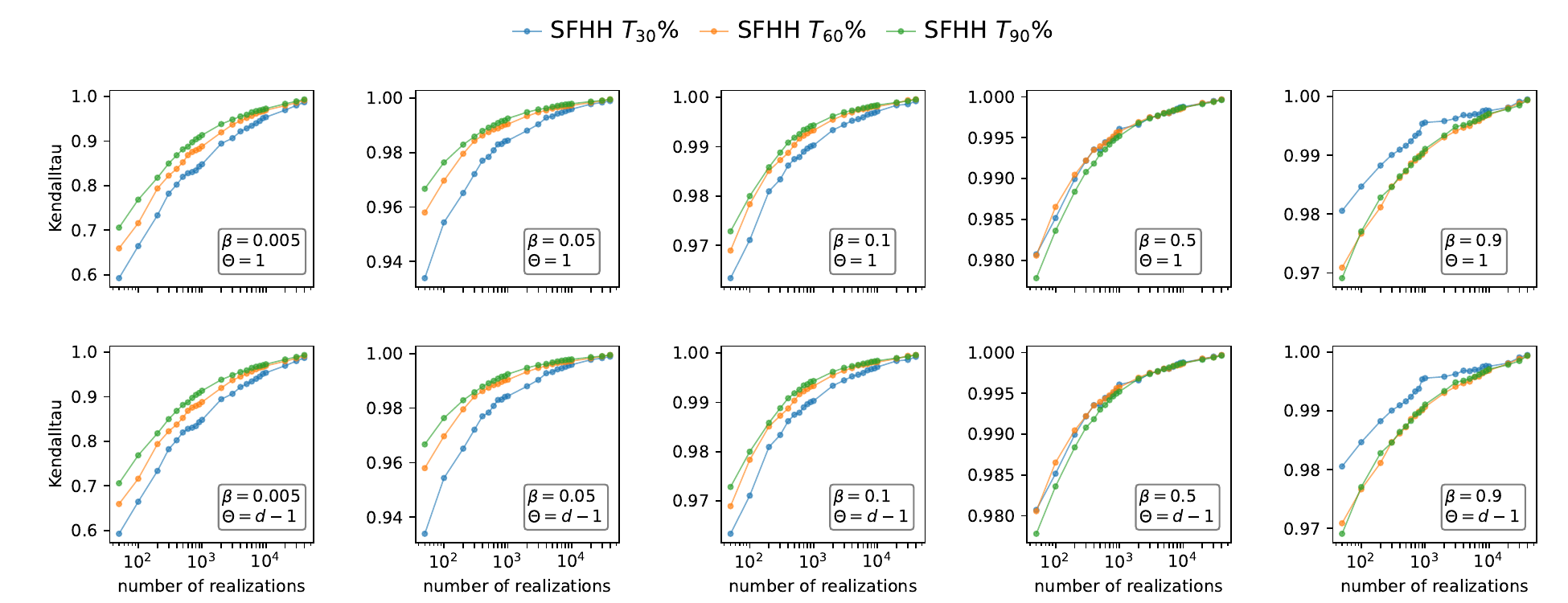}
\caption{The Kendall correlation between the weights in the backbones resulting from $R$ realizations and $50000$ realizations, as a function of $R$ for order-3 hyperlinks, in \textit{SFHH} dataset with different observation time windows.}
\label{sfig:convergence_SFHH_order3}
\end{figure*}

Figure \ref{sfig:convergence_all_order2} and Figure \ref{sfig:convergence_all_order3} show that ss the number of realizations $R$ grows, the Kendall correlation $\tau(R)$ increases quickly and approaches one, suggesting that the ranking of hyperlinks by their weights in the backbone gradually becomes stable. This supports our choice of $R=50000$ realizations. Similar trends are observed when changing the observation time window, as shown in Figure \ref{sfig:convergence_SFHH_order2} and Figure \ref{sfig:convergence_SFHH_order3}.

\clearpage
\section{supplementary results for \textit{SFHH} dataset}
\begin{figure*}[h]
\centering
\includegraphics[scale=0.9]{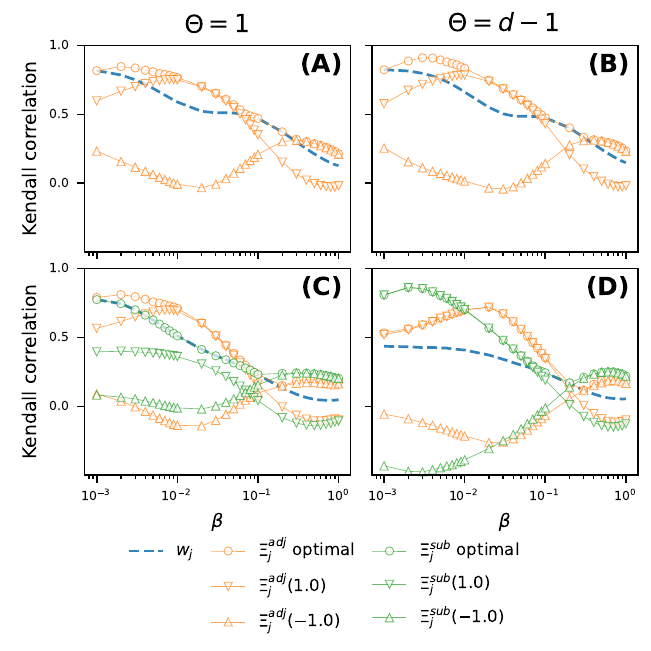}
\caption{The \textit{SFHH} dataset with the observation time window $[0, T_{90\%})$. The Kendall rank correlation between the metrics $w_j$, $\Xi_j^{adj}(1)$, $\Xi_j^{adj}(-1)$,  $\Xi_j^{sub}(1)$ and $\Xi_j^{sub}(-1)$, and the weight $w_j^B$ of a hyperlink, in comparison to the optimal Kendall correlation between metrics, $\Xi_j^{adj}$, $\Xi_j^{sub}$, and the weight $w_j^B$. Results are shown for dyadic hyperlinks (A-B) and triadic hyperlinks (C-D). Two columns corresponds to $\Theta=1$ and $\Theta=d-1$, respectively.}
\label{sfig:influence_parameters_subnet1}
\end{figure*}

\begin{figure*}[h]
\centering
\includegraphics[scale=0.8]{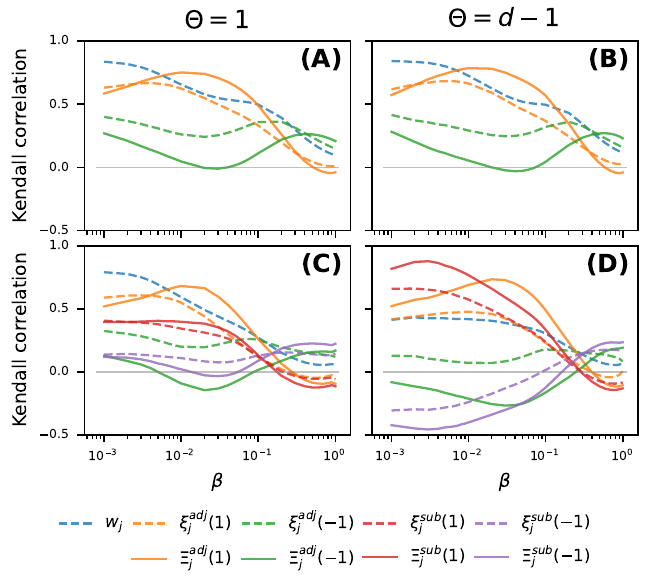}
\caption{The Kendall correlation between a centrality metric with the scaling parameter $\alpha\in\{0, 1, -1\}$ and the weight $w_j^B$ of a hyperlink in the backbone $B$, as a function of infection probability $\beta$, in \textit{SFHH} dataset with the time observation time window $[0, T_{60\%}]$. Results are shown for dyadic hyperlinks (A-B) and triadic hyperlinks (C-D). Two columns corresponds to $\Theta=1$ and $\Theta=d-1$, respectively.}
\label{sfig:influence_parameters_subnet1}
\end{figure*}

\begin{figure*}[h]
\centering
\includegraphics[scale=0.8]{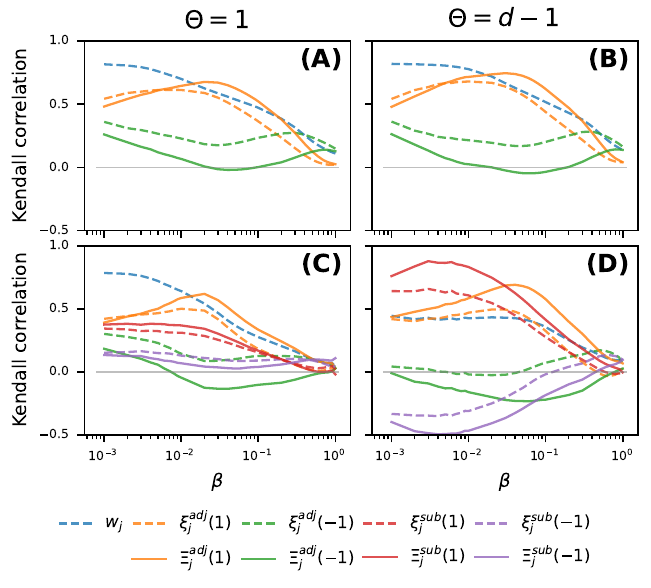}
\caption{The Kendall correlation between a centrality metric with the scaling parameter $\alpha\in\{0, 1, -1\}$ and the weight $w_j^B$ of a hyperlink in the backbone $B$, as a function of infection probability $\beta$, in \textit{SFHH} dataset with the time observation time window $[0, T_{30\%}]$. Results are shown for dyadic hyperlinks (A-B) and triadic hyperlinks (C-D). Two columns corresponds to $\Theta=1$ and $\Theta=d-1$, respectively.}
\label{sfig:influence_parameters_subnet1}
\end{figure*}

\begin{figure*}[h]
\centering
\includegraphics[scale=0.8]{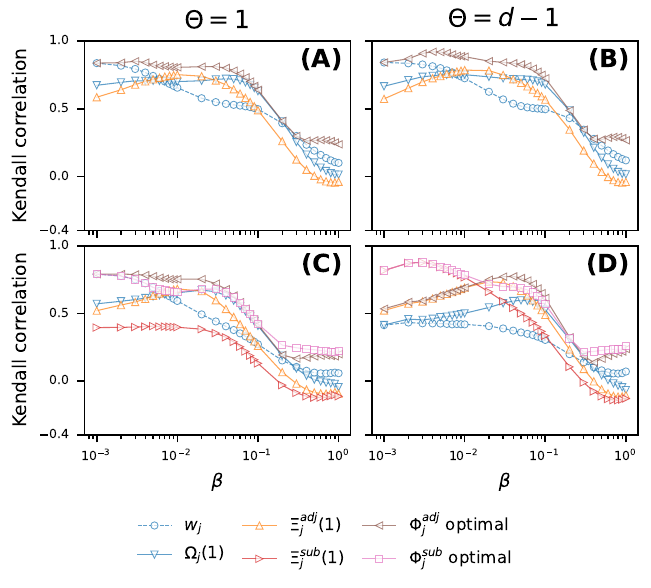}
\caption{The Kendall rank correlation between each of the four metrics $w_j$, $\Omega_j(1)$, $\Xi_j^{adj}$ and $\Xi_j^{sub}$ and the weight $w_j^B$ of a hyperlink in the backbone, in comparison to the optimal Kendall correlation between each combined metric and the weight $w_j^B$, in \textit{SFHH} dataset with the time observation time window $[0, T_{60\%}]$. Results are shown for dyadic hyperlinks (A-B) and triadic hyperlinks (C-D). Two columns corresponds to $\Theta=1$ and $\Theta=d-1$, respectively.}
\label{sfig:influence_parameters_subnet0}
\end{figure*}

\begin{figure*}[h]
\centering
\includegraphics[scale=0.8]{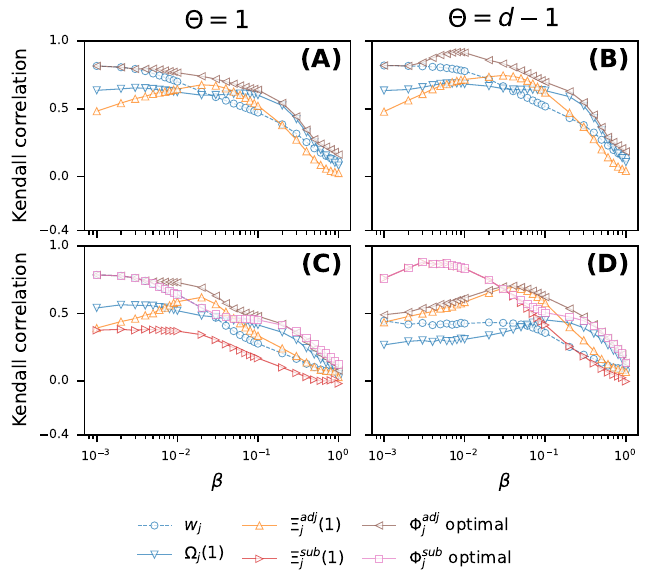}
\caption{The Kendall rank correlation between each of the four metrics $w_j$, $\Omega_j(1)$, $\Xi_j^{adj}$ and $\Xi_j^{sub}$ and the weight $w_j^B$ of a hyperlink in the backbone, in comparison to the optimal Kendall correlation between each combined metric and the weight $w_j^B$, in \textit{SFHH} dataset with the time observation time window $[0, T_{30\%}]$. Results are shown for dyadic hyperlinks (A-B) and triadic hyperlinks (C-D). Two columns corresponds to $\Theta=1$ and $\Theta=d-1$, respectively.}
\label{sfig:influence_parameters_subnet1}
\end{figure*}

\clearpage
\section{Results for other datasets}

\begin{figure*}[h]
\centering
\includegraphics[scale=0.8]{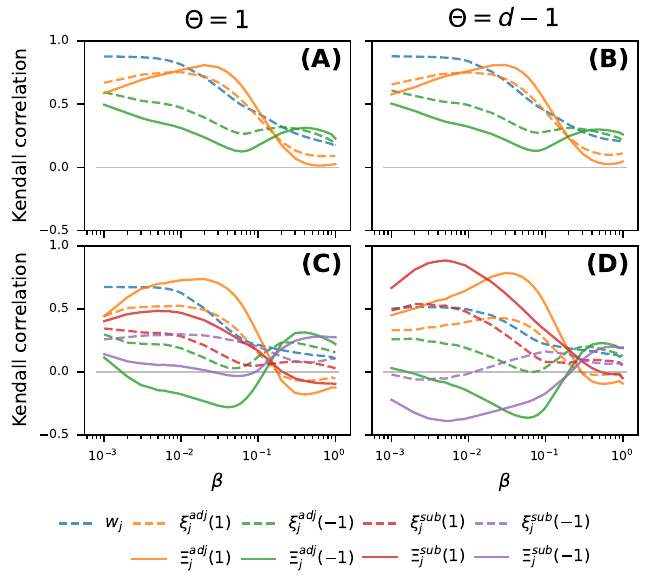}
\caption{The \textit{infectious} dataset. The Kendall correlation between a centrality metric with the scaling parameter $\alpha\in\{0, 1, -1\}$ and the weight $w_j^B$ of a hyperlink in the backbone $B$, as a function of infection probability $\beta$. Results are shown for dyadic hyperlinks (A-B) and triadic hyperlinks (C-D). Two columns corresponds to $\Theta=1$ and $\Theta=d-1$, respectively.}
\label{sfig:influence_parameters_subnet1}
\end{figure*}

\begin{figure*}[h]
\centering
\includegraphics[scale=0.8]{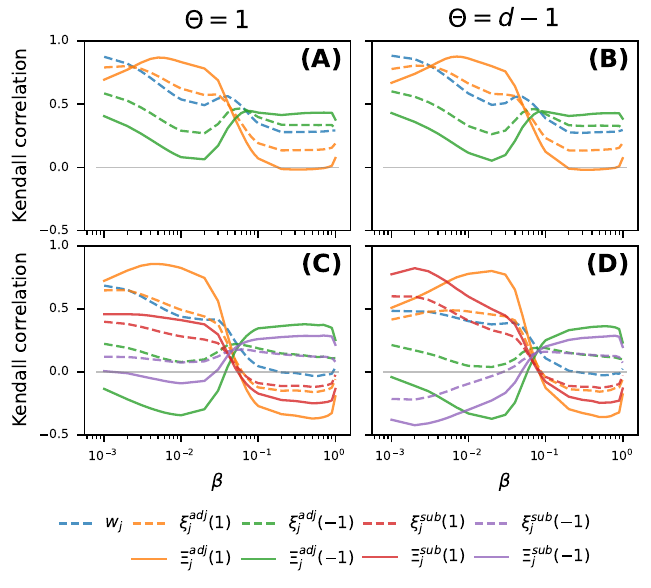}
\caption{The \textit{primaryschool} dataset. The Kendall correlation between a centrality metric with the scaling parameter $\alpha\in\{0, 1, -1\}$ and the weight $w_j^B$ of a hyperlink in the backbone $B$, as a function of infection probability $\beta$. Results are shown for dyadic hyperlinks (A-B) and triadic hyperlinks (C-D). Two columns corresponds to $\Theta=1$ and $\Theta=d-1$, respectively.}
\label{sfig:influence_parameters_subnet1}
\end{figure*}

\begin{figure*}[h]
\centering
\includegraphics[scale=0.8]{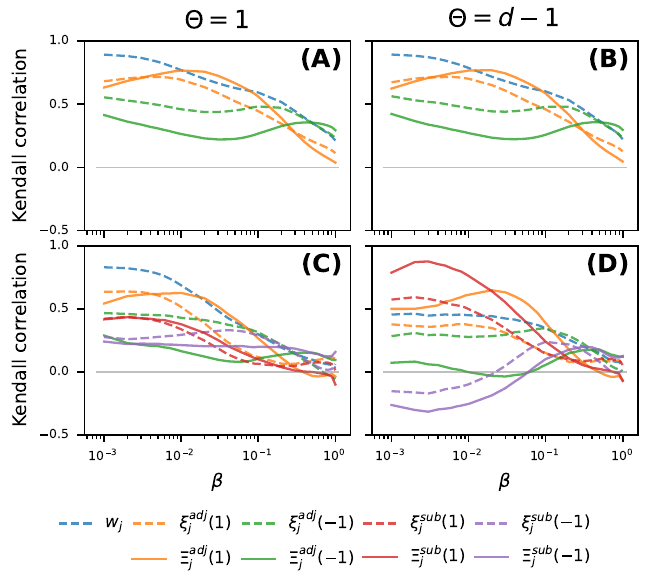}
\caption{The \textit{highschool2012} dataset. The Kendall correlation between a centrality metric with the scaling parameter $\alpha\in\{0, 1, -1\}$ and the weight $w_j^B$ of a hyperlink in the backbone $B$, as a function of infection probability $\beta$. Results are shown for dyadic hyperlinks (A-B) and triadic hyperlinks (C-D). Two columns corresponds to $\Theta=1$ and $\Theta=d-1$, respectively.}
\label{sfig:influence_parameters_subnet1}
\end{figure*}

\begin{figure*}[h]
\centering
\includegraphics[scale=0.8]{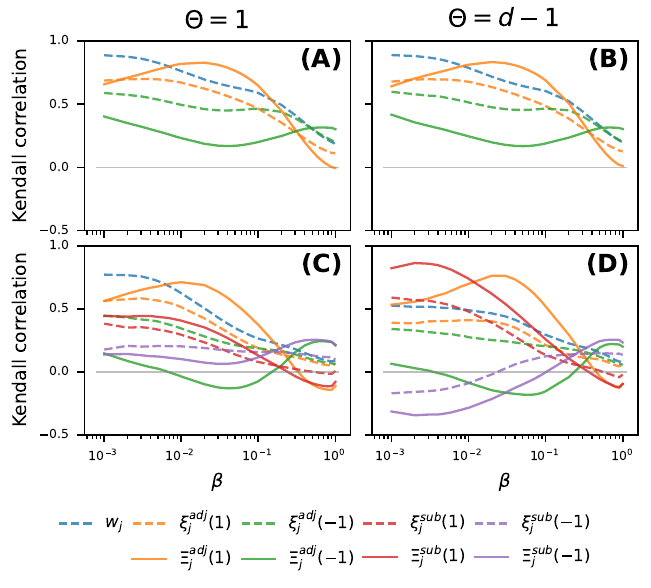}
\caption{The \textit{highschool2013} dataset. The Kendall correlation between a centrality metric with the scaling parameter $\alpha\in\{0, 1, -1\}$ and the weight $w_j^B$ of a hyperlink in the backbone $B$, as a function of infection probability $\beta$. Results are shown for dyadic hyperlinks (A-B) and triadic hyperlinks (C-D). Two columns corresponds to $\Theta=1$ and $\Theta=d-1$, respectively.}
\label{sfig:influence_parameters_subnet1}
\end{figure*}

\begin{figure*}[h]
\centering
\includegraphics[scale=0.8]{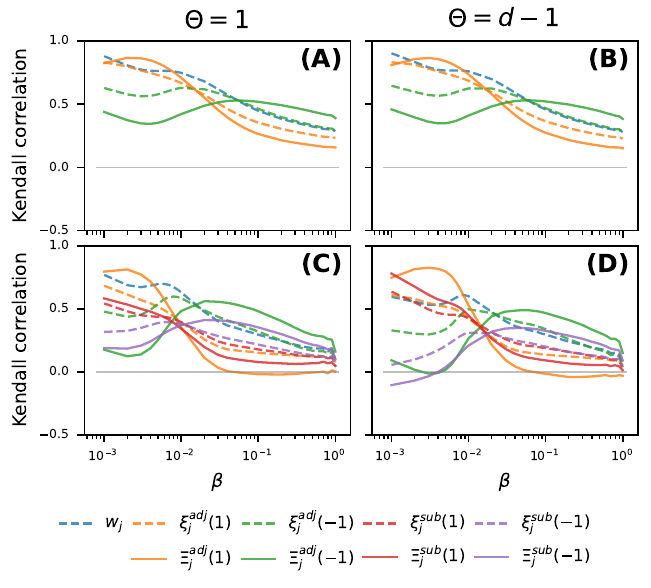}
\caption{The \textit{hospital} dataset. The Kendall correlation between a centrality metric with the scaling parameter $\alpha\in\{0, 1, -1\}$ and the weight $w_j^B$ of a hyperlink in the backbone $B$, as a function of infection probability $\beta$. Results are shown for dyadic hyperlinks (A-B) and triadic hyperlinks (C-D). Two columns corresponds to $\Theta=1$ and $\Theta=d-1$, respectively.}
\label{sfig:influence_parameters_subnet1}
\end{figure*}

\begin{figure*}[h]
\centering
\includegraphics[scale=0.8]{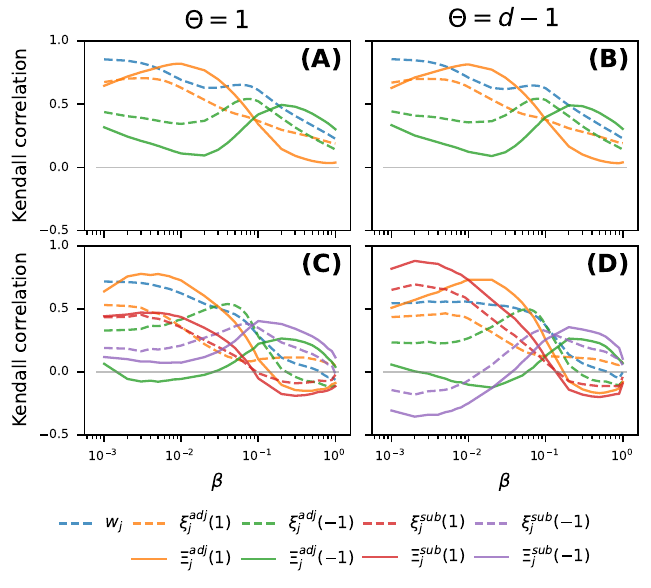}
\caption{The \textit{ht09} dataset. The Kendall correlation between a centrality metric with the scaling parameter $\alpha\in\{0, 1, -1\}$ and the weight $w_j^B$ of a hyperlink in the backbone $B$, as a function of infection probability $\beta$. Results are shown for dyadic hyperlinks (A-B) and triadic hyperlinks (C-D). Two columns corresponds to $\Theta=1$ and $\Theta=d-1$, respectively.}
\label{sfig:influence_parameters_subnet1}
\end{figure*}

\begin{figure*}[h]
\centering
\includegraphics[scale=0.8]{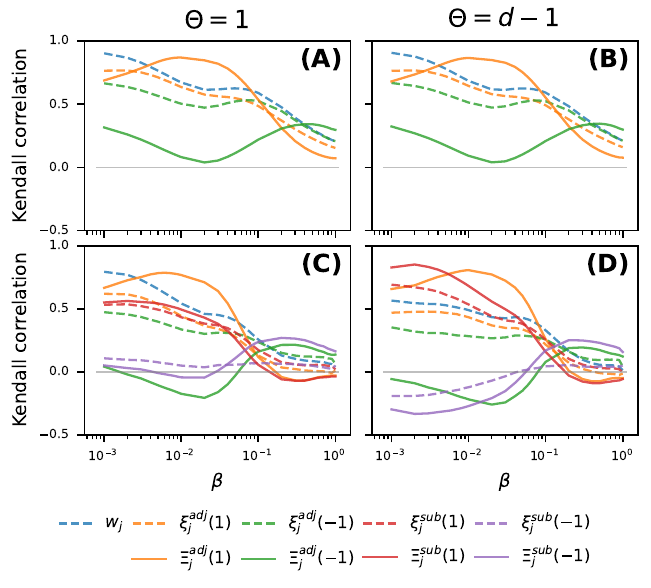}
\caption{The \textit{workplace15} dataset. The Kendall correlation between a centrality metric with the scaling parameter $\alpha\in\{0, 1, -1\}$ and the weight $w_j^B$ of a hyperlink in the backbone $B$, as a function of infection probability $\beta$. Results are shown for dyadic hyperlinks (A-B) and triadic hyperlinks (C-D). Two columns corresponds to $\Theta=1$ and $\Theta=d-1$, respectively.}
\label{sfig:influence_parameters_subnet1}
\end{figure*}

\begin{figure*}[h]
\centering
\includegraphics[scale=0.8]{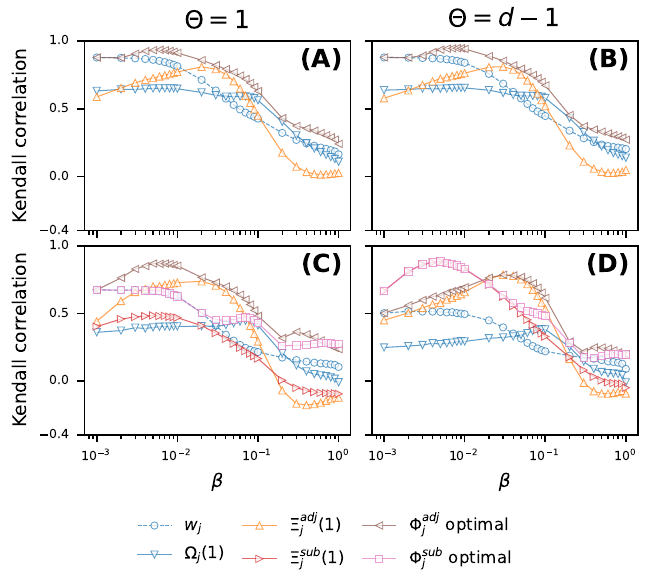}
\caption{The \textit{infectious} dataset. The Kendall rank correlation between each of the four metrics $w_j$, $\Omega_j(1)$, $\Xi_j^{adj}$ and $\Xi_j^{sub}$ and the weight $w_j^B$ of a hyperlink in the backbone, in comparison to the optimal Kendall correlation between each combined metric and the weight $w_j^B$. Results are shown for dyadic hyperlinks (A-B) and triadic hyperlinks (C-D). Two columns corresponds to $\Theta=1$ and $\Theta=d-1$, respectively.}
\label{sfig:influence_parameters_subnet1}
\end{figure*}

\begin{figure*}[h]
\centering
\includegraphics[scale=0.8]{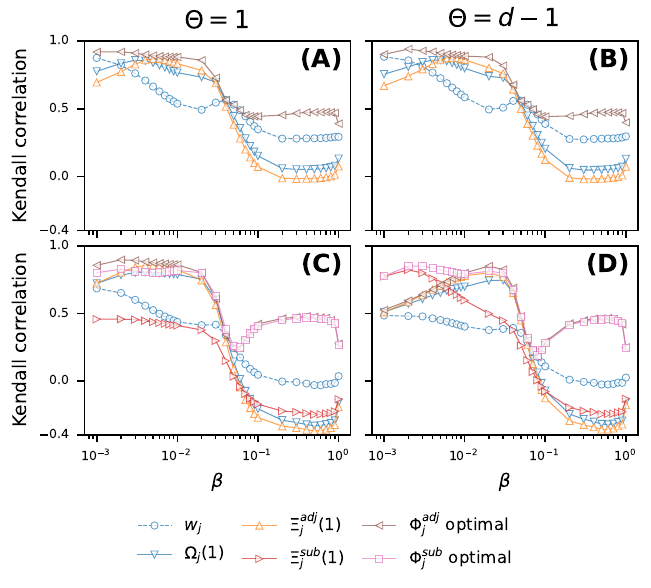}
\caption{The \textit{primaryschool} dataset. The Kendall rank correlation between each of the four metrics $w_j$, $\Omega_j(1)$, $\Xi_j^{adj}$ and $\Xi_j^{sub}$ and the weight $w_j^B$ of a hyperlink in the backbone, in comparison to the optimal Kendall correlation between each combined metric and the weight $w_j^B$. Results are shown for dyadic hyperlinks (A-B) and triadic hyperlinks (C-D). Two columns corresponds to $\Theta=1$ and $\Theta=d-1$, respectively.}
\label{sfig:influence_parameters_subnet1}
\end{figure*}

\begin{figure*}[h]
\centering
\includegraphics[scale=0.8]{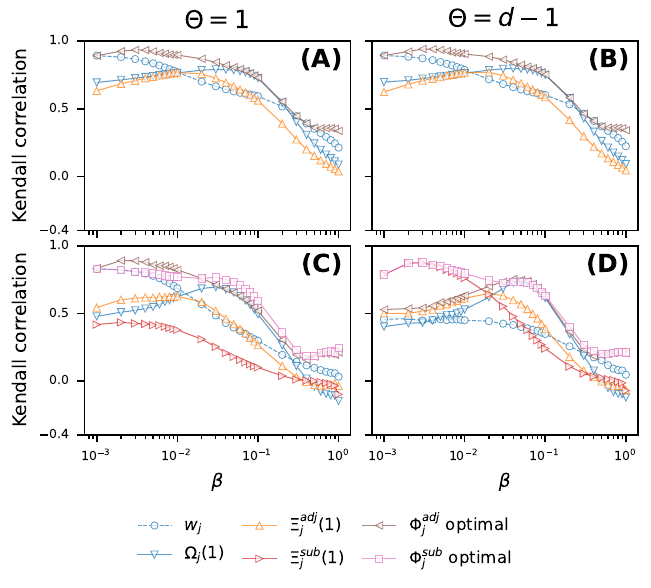}
\caption{The \textit{highschool2012} dataset. The Kendall rank correlation between each of the four metrics $w_j$, $\Omega_j(1)$, $\Xi_j^{adj}$ and $\Xi_j^{sub}$ and the weight $w_j^B$ of a hyperlink in the backbone, in comparison to the optimal Kendall correlation between each combined metric and the weight $w_j^B$. Results are shown for dyadic hyperlinks (A-B) and triadic hyperlinks (C-D). Two columns corresponds to $\Theta=1$ and $\Theta=d-1$, respectively.}
\label{sfig:influence_parameters_subnet1}
\end{figure*}

\begin{figure*}[h]
\centering
\includegraphics[scale=0.8]{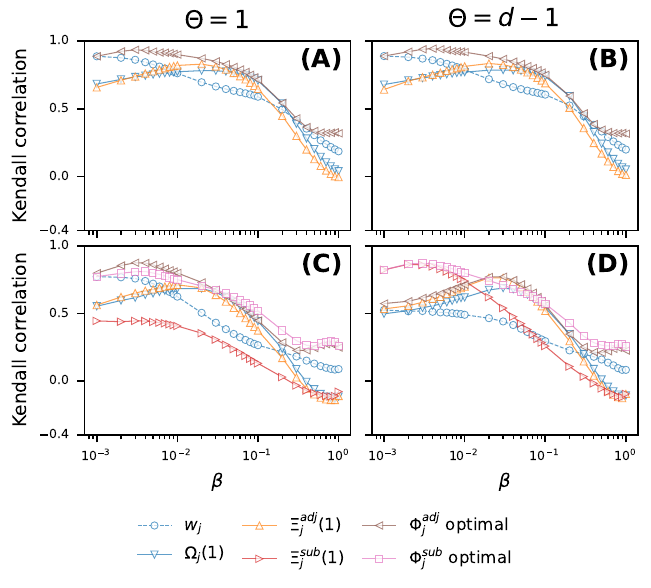}
\caption{The \textit{highschool2013} dataset. The Kendall rank correlation between each of the four metrics $w_j$, $\Omega_j(1)$, $\Xi_j^{adj}$ and $\Xi_j^{sub}$ and the weight $w_j^B$ of a hyperlink in the backbone, in comparison to the optimal Kendall correlation between each combined metric and the weight $w_j^B$. Results are shown for dyadic hyperlinks (A-B) and triadic hyperlinks (C-D). Two columns corresponds to $\Theta=1$ and $\Theta=d-1$, respectively.}
\label{sfig:influence_parameters_subnet1}
\end{figure*}

\begin{figure*}[h]
\centering
\includegraphics[scale=0.8]{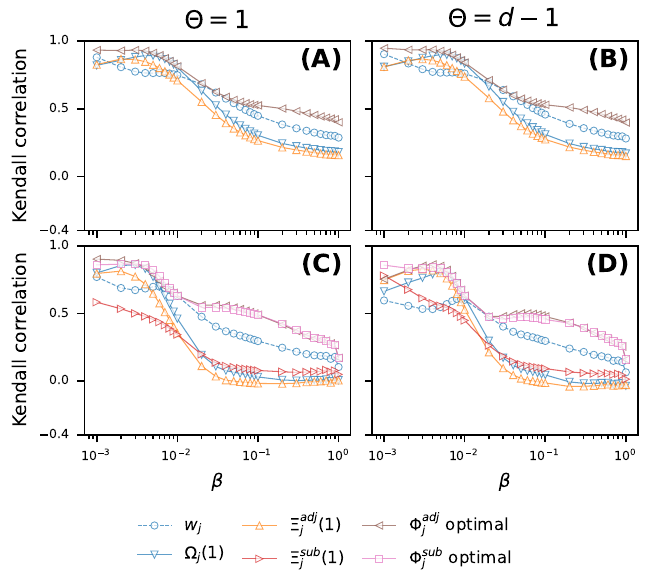}
\caption{The \textit{hospital} dataset. The Kendall rank correlation between each of the four metrics $w_j$, $\Omega_j(1)$, $\Xi_j^{adj}$ and $\Xi_j^{sub}$ and the weight $w_j^B$ of a hyperlink in the backbone, in comparison to the optimal Kendall correlation between each combined metric and the weight $w_j^B$. Results are shown for dyadic hyperlinks (A-B) and triadic hyperlinks (C-D). Two columns corresponds to $\Theta=1$ and $\Theta=d-1$, respectively.}
\label{sfig:influence_parameters_subnet1}
\end{figure*}

\begin{figure*}[h]
\centering
\includegraphics[scale=0.8]{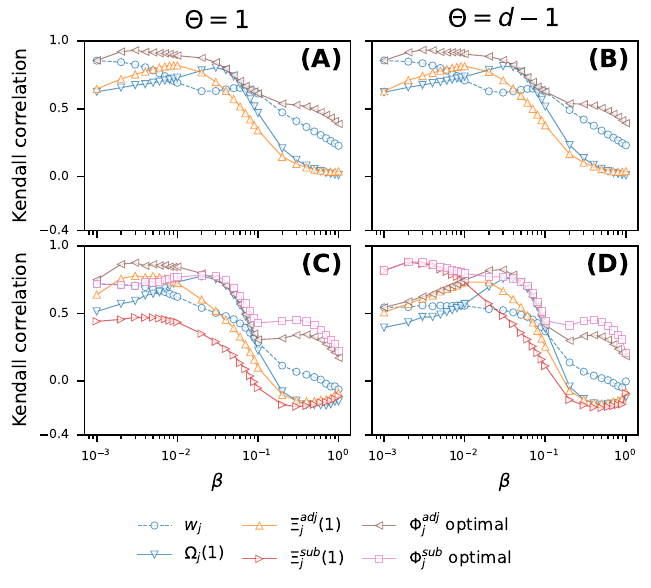}
\caption{The \textit{ht09} dataset. The Kendall rank correlation between each of the four metrics $w_j$, $\Omega_j(1)$, $\Xi_j^{adj}$ and $\Xi_j^{sub}$ and the weight $w_j^B$ of a hyperlink in the backbone, in comparison to the optimal Kendall correlation between each combined metric and the weight $w_j^B$. Results are shown for dyadic hyperlinks (A-B) and triadic hyperlinks (C-D). Two columns corresponds to $\Theta=1$ and $\Theta=d-1$, respectively.}
\label{sfig:influence_parameters_subnet1}
\end{figure*}

\begin{figure*}[h]
\centering
\includegraphics[scale=0.8]{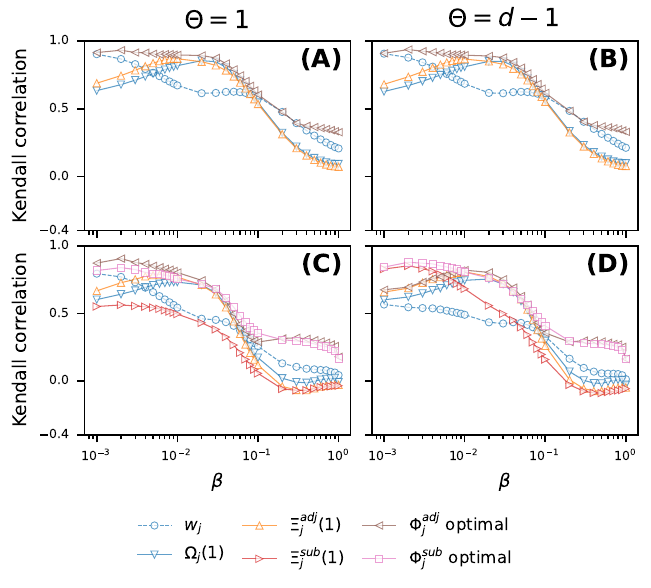}
\caption{The \textit{workplace15} dataset. The Kendall rank correlation between each of the four metrics $w_j$, $\Omega_j(1)$, $\Xi_j^{adj}$ and $\Xi_j^{sub}$ and the weight $w_j^B$ of a hyperlink in the backbone, in comparison to the optimal Kendall correlation between each combined metric and the weight $w_j^B$. Results are shown for dyadic hyperlinks (A-B) and triadic hyperlinks (C-D). Two columns corresponds to $\Theta=1$ and $\Theta=d-1$, respectively.}
\label{sfig:influence_parameters_subnet1}
\end{figure*}

\end{document}